\documentclass[]{aastex}

\usepackage{emulateapj5}
\usepackage{onecolfloat}
\usepackage{graphicx} 
\usepackage{fancyheadings} 
\usepackage{ulem}
\usepackage{rotating}
\usepackage{lscape}

\newcommand{\etal}{et al.\ }

\newcommand{\lya}{Ly$\alpha$ }
\newcommand{\kms}{km~s$^{-1}$ }

\newcommand{\ohf}{{1 \over 2}}

\newcommand{\perd}{\;\;\; .}
\newcommand{\cmma}{\;\;\; ,}
\newcommand{\semic}{\;\;\; ;}

\newcommand{\Nperp}{N_\perp(0)}
\newcommand{\NL}{$2{\times}10^{20}$cm$^{-2}$}

\newcommand{\dv}{$\Delta v$}
\newcommand{\dvm}{$\delta v$}

\newcommand{\dciv}{$\Delta v_{\rm C IV}$}
\newcommand{\dlow}{$\Delta v_{\rm low}$}

\newcommand{\eslash}{$\rm \acute e$}
\newcommand{\lcdm}{$\Lambda$CDM}

\newcommand{\vc}{$V_{200}$}
\newcommand{\vm}{$V_{max}$}
\newcommand{\vcut}{$V_{cut}$}
\newcommand{\vrot}{$V_{rot}(r)$}
\newcommand{\nh}{$N$(H I)}

\newcommand{\rc}{$r_{cool}$}
\newcommand{\rd}{$R_{d}$}
\newcommand{\rv}{$r_{200}$}
\newcommand{\rcr}{$r_{cross}$}
\newcommand{\rmin}{$r_{min}$}
\newcommand{\frat}{$f_{ratio}$}

\begin{document}

\twocolumn[%
\submitted{accepted by the Astrophysical Journal August 5, 2000}

\title{IONIZED GAS IN DAMPED \lya\ PROTOGALAXIES: II. COMPARISON BETWEEN MODELS
AND THE KINEMATIC DATA }

\author{ ARTHUR M. WOLFE\altaffilmark{1}\\
Department of Physics, and Center for Astrophysics and Space Sciences; \\
University of California, San
Diego; \\
C--0424; La Jolla; CA 92093\\
{\bf awolfe@ucsd.edu}}

\author{and}

\author{ JASON X. PROCHASKA\altaffilmark{1}\\
The Observatories of the Carnegie Institute of Washington; \\
813 Santa Barbara St.\\
Pasadena, CA; 91101\\
{\bf xavier@ociw.edu}}

\begin{abstract}

We test semi-analytic models for galaxy formation with
accurate kinematic data of damped {\lya}
protogalaxies presented in a companion
paper (Wolfe \& Prochaska 2000). The models envisage
centrifugally
supported exponential disks at the centers of dark matter halos
which are filled with ionized gas undergoing radial infall to the disks. The
halo masses are drawn from cross-section weighted
mass distributions predicted by CDM cosmogonies,
or  by the
null hypothesis that the dark matter mass distribution
has not evolved since
$z$ $\sim$ 3 (i.e., the TF models). In our models
C IV absorption lines detected in damped {\lya} protogalaxies
arise in infalling ionized clouds
while the low ion absorption lines arise from neutral
gas in the disks. Using Monte Carlo methods we find
(a) The CDM models are incompatible with the low ion statistics
at more than 99$\%$ confidence whereas some TF models
cannot be ruled out at more than 88$\%$ confidence.
(b) Both CDM and TF models are in general agreement with the
observed distribution of C IV velocity widths. (c) The
CDM models generate differences between the mean velocities of
C IV and low ion profiles that are compatible with the data, while the
TF model produces differences in the means  that are too large. (d) Both CDM and
TF models produce  ratios of C IV to low ion velocity
widths that are too large.  (e) Neither CDM
nor TF models generate C IV versus low ion cross-correlation
functions compatible with the data.

While it is possible to select model parameters resulting in
agreement between the models and the data, the fundamental problem
is that the disk-halo configuration assumed in both cosmogonies
does not produce significant overlap in velocity space between
C IV and low ion velocity profiles. We conjecture that including angular momentum
of the infalling clouds will increase the overlap between C IV and low
ion profiles.

 \end{abstract}

]
\altaffiltext{1}{Visiting Astronomer, W.M. Keck Telescope.
The Keck Observatory is a joint facility of the University
of California and the California Institute of Technology.}

\pagestyle{fancyplain}
\lhead[\fancyplain{}{\thepage}]{\fancyplain{}{WOLFE \& PROCHASKA}}
\rhead[\fancyplain{}{IONIZED GAS IN DAMPED {\lya} PROTOGALAXIES II}]{\fancyplain{}{\thepage}}
\setlength{\headrulewidth=0pt}
\cfoot{}

\section{INTRODUCTION}

This is the second of two papers which discuss ionized gas
in damped {\lya} systems. In Paper I (Wolfe \& Prochaska 2000)
we presented velocity profiles drawn from a sample of 35 damped
{\lya} systems for the high ions C IV and Si IV
and the intermediate ion Al III.
Comparison among these profiles and with profiles previously obtained
for low ions such as Fe II showed the damped {\lya} systems to
consist of two distinct kinematic subsystems: a low ion subsystem
composed of low and intermediate ions and a high ion subsystem
containing only high ions. The evidence distinguishing between
the kinematic subsystems is robust and stems from a battery of
tests comparing distributions of various test statistics. It
also stems from differences between the C IV versus low ion or C IV
versus Al III
cross-correlation functions on the one hand, and the C IV versus Si IV or Al III
versus low ion cross correlation functions on the other. Whereas the latter
have high amplitude and small velocity width, the former have lower
amplitude and wider velocity width. This is because velocity profiles
of ions
arising in
the same kinematic subsystem comprise narrow velocity components
that line up in velocity space, whereas velocity components arising
from ions in different kinematic subsystems are misaligned in
velocity space. However, the existence of a statistically
significant C IV versus
low ion cross-correlation function suggests the two subsystems
are interrelated. This is also indicated by a systematic effect
in which the C IV profile velocity widths {\dciv} are greater than or equal
to the low ion profile velocity widths {\dlow} in 29 out of 32 systems.

In Paper I we claimed these phenomena indicate
that the two subsystems are located in the same gravitational potential
well.
In this paper we shall expand on this idea with specific models.
The models assume that the low ion subsystems are
centrifugally supported disks of neutral gas
located at the centers
of dark matter halos (see Mo {\etal} 1998; hereafter referred to
as MMW), whereas the high ion subsystems comprise photoionized
clouds undergoing infall from a gaseous halo to the disk.
That is, we assume the dark matter halos contain
hot virialized gas in pressure equilibrium with the photoionized
clouds. The hot gas undergoes a subsonic cooling flow toward the
disk while the denser clouds infall at velocities approaching
free fall
(Mo \& Miralda-Escud{\eslash} 1996; hereafter
referred to as MM).
The models are set in a cosmological context by
computing
the mass  distribution and other properties
of the dark matter halos
using Press-Schecter theory and the CDM cosmogonies adopted
by semi-analytic models for galaxy formation.
We also consider the null hypothesis
that galaxies at $z$ $\approx$ 3 have the same dynamical properties
of nearby galaxies (this model, hereafter referred
to as the TF model, is defined in $\S$ 2.4.2).

We test the models using a
Monte
Carlo technique for computing absorption spectra arising when
sightlines that randomly penetrate intervening disks also
intercept ionized clouds in the halo. $\S$ 2 presents models for the
neutral gas. We discuss properties of the disk, the cosmological
framework, and Monte Carlo techniques.
In
$\S$ 3 we discuss models for the ionized gas. We derive expressions
for the structure and kinematics of the two-phase halo gas.
Expressions for  C IV column densities of the clouds are derived.
In $\S$ 4 we give results of the Monte Carlo simulations for the low ion
gas for both CDM and TF model. Results of Monte Carlo simulations for
the ionized gas are given in $\S$ 5. Here we also  consider tests
of correlations between the kinematics of the high ions and low ions.
In $\S$ 6 we investigate how the results of $\S$ 5 are affected
by changes in some key parameters such as central column density and
low end cutoff  to the input dark-matter halo mass distribution.
Concluding remarks are given in $\S$ 7.

\section{MODELS OF THE NEUTRAL GAS}
\label{mdls}

\subsection{Cosmological Framework}

To place the model in a cosmological context
we assume bound dark-matter halos evolve from
linear density contrasts, $\delta(x,t)$$\equiv$$\delta \rho/\rho$,
according to gravitational instability
theory for Friedmann cosmologies \citep{pbl1980}.
We consider adiabatic CDM models (ACDM) in which
$\delta_{k}$, the Fourier transforms of $\delta(x,t)$,
are Gaussian distributed with variance given by
$P(k)$,
the power spectrum at the epoch of radiation and matter
equality. We also consider isocurvature CDM models (ICDM)
in which the $\delta_{k}$ are not Gaussian distributed \citep{pbl99b}.
The $\delta_{k}$, or more specifically the rms
density contrasts in spheres with mass scale $M$, i.e., $\Delta_{M}$,
grow with time until
they go non-linear and collapse. According to the spherical collapse
model, this occurs when the $\Delta_{M}$ predicted
by linear theory equals $\delta_{c}$ = 1.68.
To compute $n(M,z)dM$, the density of bound halos in the mass interval
($M$, $M$$+$$dM$), we follow previous authors (e.g. MMW) who
used the Press-Schecter formalism in the case of ACDM. In Appendix A
we derive an expression for $n(M,z)dM$ in the case of ICDM.

We shall also test the null hypothesis that little or no
evolution of galaxies has occurred since the epochs of
damped {\lya} absorption; that is, an hypothesis assuming
current disks, with higher ratios of gas to stars,  were in place
at  $z$ $>$ 3. This scenario resembles the semi-analytic models
in that we assume centrifugally supported disks reside
at the centers of dark matter halos filled with hot gas
at the virial temperatures. However,
we do not assume a CDM power spectrum nor the Press-Schecter formalism
to compute
the mass distribution of halos. Rather, we assume
(a) the luminosity
function of galaxies in the redshift range of the damped {\lya}
sample in paper I, i.e.,
$z$ = [2,5],
is given
by the Schecter function determined from
nearby galaxies (e.g. Loveday {\etal} 1999),
(b) galaxies
at these redshifts
obey the same Tully-Fisher relationship between luminosity and
rotation speed as at $z$ = 0, and (c) a correlation between disk
radial scale-length and disk rotation speed exists.
This model, hereafter referred to as TF, is an extension of the rapidly rotating disk
model suggested by Prochaska \& Wolfe (1997 and 1998; hereafter PW1 and PW2).
We describe this model in more detail in the following sections
\footnote{We could have
adopted luminosity functions that are {\em measured} in this redshift interval. The most
accurate determinations are
for the Lyman-break galaxies (Steidel {\etal} 1999). However, we
rejected this procedure because
the
rotation speeds of these objects have not been measured, nor
is it known whether or not they contain rotating disks.
The strong clustering exhibited by the Lyman-break galaxies suggests
otherwise. Nevertheless, we briefly discuss this possibility
in  $\S$ 6.2}

Throughout the paper we shall intercompare results for the TF model
and four CDM cosmogonies.
The cosmological settings of
the CDM models are specified by (i) the current total
matter density,  $\Omega_{M}$, (ii)
the cosmological constant, $\Omega_{\Lambda}$, (iii) the Hubble constant, $h$, where
$h$ = $H_{0}$/ 100 {\kms} Mpc$^{-1}$, (iv) $\sigma_{8}$, the rms linear density contrast at
$z$ = 0 in spheres of radius 8$h^{-1}$ Mpc,
and (v)  $n$, the power-law index for the power spectrum, in cases where
$P(k)$ $\propto$ $k^{n}$. The values of the parameters are given in
Table~\ref{tab:modparm}.
The SCDM, $\Lambda$CDM, and OCDM are normal ACDM models. In all three
cases, $P(k)$ is given by the \cite{bard86} expression (which is not
a power law). In the
ICDM model we follow Peebles by assuming $P(k)$ $\propto$ $k^{n}$ and
$n$ = $-$1.8. The cosmological parameters
specifying the TF model are also given in Table~\ref{tab:modparm}.

\subsection{Disk Models}

Most semi-analytic models assume the neutral gas causing
damped {\lya} absorption is confined to centrifugally supported disks
at the centers of dark-matter halos (e.g., Kauffmann 1996;
MMW). The spherical collapse model  is used
to define the limiting virial radius as $r_{200}$,  the radius
within which the mean density of dark matter equals 200$\rho_{crit}(z)$ where
$\rho_{crit}(z)$ is the critical density of the universe at redshift $z$. Thus,
{\vc}, the circular velocity at {\rv}, is related to {\rv} and
$M$, the halo mass within {\rv}, by

\begin{equation}
V_{200} = [10GH(z)M]^{1/3}; \  r_{200} =  {V_{200} \over {10H(z)}}
\cmma
\end{equation}

\noindent where
$H(z)$ is the Hubble parameter, ($a^{-1}da/dt$), at $z$.
MMW also define $m_{d}$ and $j_{d}$
as the
fractions of halo mass, $M$, and angular momentum, $J$,
in disk baryons.
Assuming the halos to be singular isothermal spheres embedding
disks having exponential surface-density profiles with radial scale-lengths,
$R{_d}^{iso}$, they find that

\begin{eqnarray}
R_{d}^{iso} = {1 \over \sqrt 2}{j_{d} \over m_{d}}{\lambda}{r_{200}}  \approx 8.8 \ h^{-1}{\rm kpc}{\Bigg(}
{\lambda \over
0.05}{\Bigg)} {\nonumber} \\ {\times}{\Bigg(}{V_{200} \over {250 {\rm km s^{-1}}}}{\Bigg)}{\Bigg[}{H(z) \over H_{0}}{\Bigg]^{-1}}{\Bigg(}{j_{d} \over m_{d}}{\Bigg)}
\cmma
\end{eqnarray}

\noindent and

\begin{eqnarray}
N_{0}^{iso} = 4.8{\times}10^{22}h \ {\rm cm^{-2}} {\Bigg(}{m_{d} \over 0.05}{\Bigg)}{\Bigg(}{\lambda \over 0.
05}{\Bigg)^{-2}} {\nonumber} \\
{\times}{\Bigg(}{V_{200} \over {250 {\rm km s^{-1}}}} 
{\Bigg)}{\Bigg[}{H(z) \over H_{0}}{\Bigg]}{\Bigg(}{m_{d} \over j
_{d}}{\Bigg)^{2}}
\cmma
\end{eqnarray}

\noindent where  the spin parameter of the halo, $\lambda$ = $J|E|^{1/2}G^{-1}M^{-5/2}$,
($E$ is the total energy of the halo) and
$N_{0}$ is the central H I column density perpendicular to the disk.
The distribution of $\lambda$ is given by

\begin{equation}
p({\lambda})d{\lambda} = {1 \over {\sqrt {2\pi}}\sigma_{\lambda}}{\rm exp} {\Bigg[}
-{{\rm ln}^{2}({\lambda/<\lambda>}) \over 2\sigma_{\lambda}^{2}}{\Bigg]}{d{\lambda} \over {\lambda}}
\cmma
\end{equation}

\noindent where the mean and dispersion are determined from
numerical simulations to be $<\lambda>$ = 0.05 and
$\sigma_{\lambda}$ =0.5 (see Barnes \& Efstathiou 1987).

For the TF models we infer the  parameters of
the halo from observed properties of the disk. Thus,
we use model rotation curves to infer {\vc} from
$V_{max}$, the observed maximum rotation speed (see below). We
then use eq. (1) to obtain the mass and virial radius of the halo.
In this case we obtain the radial scalelength and central
column density of the disk by
adopting the following correlations
inferred by MMW from the spiral galaxy data of Courteau (1996; 1997):

\begin{equation}
R_{d}^{Cor} = -1.25+7.4(V_{max}/250 {\rm km s}^{-1}) \ h^{-1}  \ {\rm kpc}
\cmma
\end{equation}

\noindent

\begin{equation}
N_{0}^{Cor} = 3.6{\times}10^{22+0.45(V_{max}/250 {\rm km s}^{-1})}{\mu}^{-1} \ h \ {\rm cm}^{-2}
\cmma
\end{equation}

\noindent where
$\mu$ is the mean molecular weight of the gas.

\subsection{Rotation Curves}

\begin{figure}[ht]
\includegraphics[height=3.3in, width=3.3in]{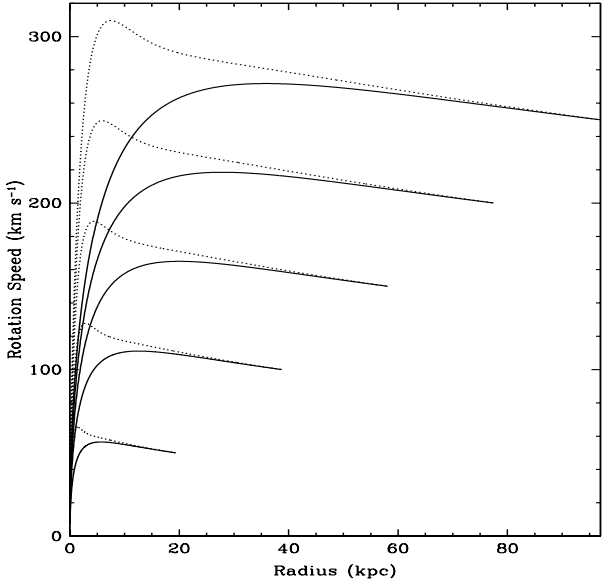}
\caption{NFW rotation curves corresponding to $z$ = 2.5, $\lambda$ =0.05,
$m$ = 0.05 and the {$\Lambda$}CDM cosmology. Solid curves are halo rotation
curves from eq. (9) corresponding to {\vc} = 50, 100, 150, 200, and 150
{\kms}. Dotted curves show rotation curves resulting from modifications
due to adiabatic contraction and disk self-gravity. Curves plotted out
to $r_{200}$}
\label{NFW_rotcurve}
\end{figure}

We next turn to the mass distribution of the halo. This is crucial
for determining both the rotation curve of the disk and the dynamics of gaseous
infall discussed in $\S$ 3. For our model we adopt the analytical fit to
the halo mass distribution found in the N-body simulations
of Navarro {\etal} (1997; hereafter NFW).
In this case the halo rotation curve, defined by $V_{rot}(r) {\equiv}  \sqrt {GM(r)/r}$, has
the following form:

\begin{equation}
V_{rot}(r) =  V_{200} {\sqrt {{c \over x} {{{\rm ln}(1+x)-x/(1+x)} \over
{\rm ln}(1+c)-c/(1+c)}}}
\semic x=r/r_{s}
\cmma
\end{equation}

\noindent where the concentration parameter, $c \equiv  r_{200}/r_{s}$, and
at $r$ = $r_{s}$,  the halo mass density
$\rho  \propto r^{-2}$.
NFW developed a self-consistent
theory in which $c$
depends on $V_{200}$ and redshift, $z$, in the sense that
at a given $z$, $c$ declines with increasing $V_{200}$, and
at a given $V_{200}$, c decreases with $z$. We
use the algorithms described in
Navarro {\etal} (1997) to compute $c$=$c(z,V_{200})$.
NFW halo rotation curves corresponding to $z$ = 2.5, $\lambda$ = 0.05, $m$ = 0.05,
and the ${\Lambda}$CDM cosmology
appear as solid curves in Figure~\ref{NFW_rotcurve}  for {\vc} = 50$-$250 {\kms}.
As expected $V_{rot}$ = $V_{max}$  at $r$ $\approx \ 2r_{s}$.
Curves with lower {\vc} appear to rise more rapidly in the interval
$r$ = [0,{\rv}] because the ratio $r_{s}$/{\rv} decreases with decreasing {\vc}.

However,
the expression for the halo {\vrot} in eq. (7) is incomplete as it
ignores the presence of the disk. Self-gravity of the disk affects the
mass distribution of the halo
through adiabatic contraction  (Blumenthal {\etal} 1984).
We used the MMW formalism to compute rotation curves due to contracted halos
and found
that {\vrot} differed from the expression in eq. (7) by less than $\approx$ 15$\%$.
Given the uncertainties in the models we shall ignore these
corrections
and use the halo mass
distribution implied by eq. (7)
when computing dynamics of {\em infall}.

On the other hand
the rotation curve of the {\em disk} can differ significantly from eq. (7)
when the disk contribution to the net potential gradient is added
to that of the contracted halo.
MMW compute the scale length  and central column density
of centrifugally supported exponential disks in adiabatically
contracted NFW halos formed by spherical collapse. In comparison
with halos modeled as singular isothermal spheres they find

\begin{equation}
R^{MMW}_{d}={R_{d}^{iso}}{f_{c}^{-1/2}}f_{R}
\cmma
N^{MMW}_{0}={N_{0}^{iso}}f_{c}f_{R}^{-2}
\perd
\end{equation}

\noindent Explicit expressions for the functions $f_{c}$ and $f_{R}$ are
given by MMW. They then use the  new disk parameters to
calculate the modified {\vrot} for disks embedded in
NFW halos.
Examples of modified {\vrot} corresponding to $\lambda$ = 0.05, $m_{d}$ = 0.05,
and $z$ = 2.5 are shown as
dotted curves in Figure~\ref{NFW_rotcurve}.
Clearly the rotation speeds sampled by sightlines traversing these model
disks lie between {\vc} and $f_{V}${\vc}, the maximum of the modified disk rotation
curve,  where $f_{V}$, a function tabulated by MMW, exceeds
unity.
Given the wide range of possible rotation
curves appropriate for model damped {\lya} systems (see PW2),
we assume that the {\em disk} rotation
curves have either of the following constant speeds:

\begin{equation}
V_{rot}(r) =V_{200}{\times} \left\{
\begin{array}{ll}
1 \   \\
f_{V} \ 
\perd
\end{array}
\right.
\end{equation}

For the TF models {\vc} is not given {\em apriori}. Rather we assume {\vrot} = {\vm}
and that {\vm} is selected from
an empirical distribution derived from the
Tully-Fisher relation (see $\S$ 2.4.2).
The radial scale length and central column densities then follow from
eqs. (5) and (6).
In this case
{\vc} = $V_{max}$ or {\vc}
=$V_{max}$/$f_{V}$.

\subsection{Monte Carlo Models}

In previous work (PW1, PW2) we tested models by
comparing predicted and empirical
distributions of the four test statistics
for the low ions (Figure 6 in Paper I).
The model distributions were computed by
a Monte Carlo technique in which
low ion
absorption profiles were produced by sightlines traversing 10000 randomly
oriented disks,
and test statistics were determined for each profile.
In PW1 the disk models were
based on the simplifying assumption of {\em identical} disks
with flat rotation curves characterized by a single rotation
speed, while in PW2 more realistic forms of {\vrot} were used
for the identical disks.
Here we extend this approach to account for the distributions
of halo masses and
spin parameters.

\subsubsection{CDM}

MMW give an expression for the  cumulative
probability that sightlines to the background QSOs intercept disks in halos with
masses exceeding that corresponding to
$V_{max}$:
the result is averaged over
$p(\lambda)$.
For the Monte Carlo model we require
the differential
expression; that is,
the probability for intercepting disks in the
spin parameter interval ($\lambda, \ \lambda + d\lambda$), circular velocity
interval ($V_{200},\ V_{200} +dV_{200}$), and redshift interval ($z,\ z  +dz$). The result is given by

\begin{eqnarray}
dP_{CDM}(z,{\lambda},V_{200}) = 
\Bigl [ dz {{\pi} \over 2} (1+z)^{3} \bigl ( -c {dt \over dz} \bigr )
\Bigr ] \nonumber \\ 
\times d \lambda p( \lambda )
 \Bigl [ R_d^{iso}(V_{200}, \lambda ,z)N_0^{iso} (V_{200},\lambda,z)
 \Bigr ]^2 \nonumber \\ \times dV_{200}n(V_{200},z)F(N_0^{iso}) 
\cmma
\end{eqnarray}

\noindent where
\begin{equation}
F(N_{0}^{iso}) = \left\{
\begin{array}{ll}
{1 \over (N_{0}^{iso})^{2}}{\Biggl[}{1 \over 2}+{\rm ln}(N_{0}^{iso}/N_{l}){\Bigl (}1 + {\rm ln}(N_{0}^{iso}/
N_{l}){\Bigr )}{\Biggr]} \ ;\\
N_{0}^{iso} > N_{l}, \\
{1 \over 2N_{l}^{2}} \ ; N_{0}^{iso} < N_{l},
\end{array}
\right.
\end{equation}

\noindent and

\begin{equation}
n(V_{200},z) = n(M,z){\Biggl |}{dM \over dV_{200}}{\Biggr |}
\perd
\end{equation}

\noindent Here $N_{l}$ = {\NL} is the threshold column density for damped {\lya}
surveys (e.g. Wolfe {\etal} 1995).

\begin{figure}
\includegraphics[height=3.3in, width=3.3in]{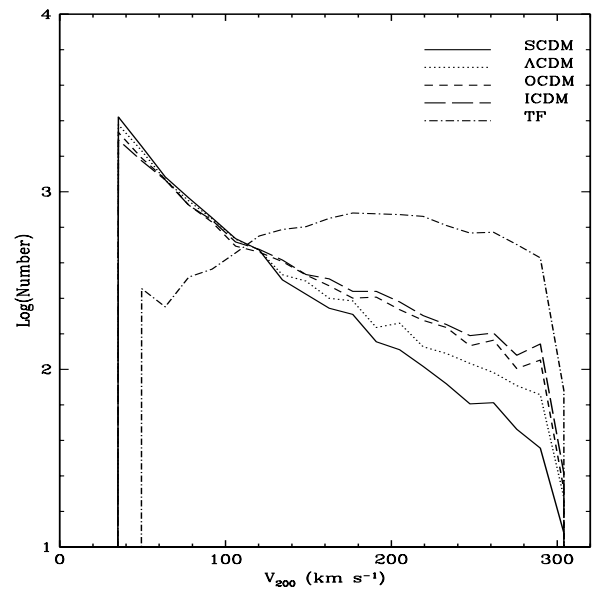}
\caption{Input circular velocity  distributions for the CDM and TF models
used in Monte Carlo simulations. Computed as described in the text
for the velocity interval (30, 300) {\kms}. The abscissa correspond
to {\vc} in CDM and {\vm} in TF models.}
\label{inputv200_bw}
\end{figure}

For given redshift intervals and cosmologies,  the function $dP_{CDM}$ describes
a surface above the ($\lambda$, $V_{200}$) plane. To form synthetic samples
of 10000 low ion profiles, we randomly draw ($\lambda$, $V_{200}$) pairs according
to the height of the surface above the plane. To assure compliance with the
damped {\lya} surveys, sightlines resulting in observed H I column densities
less than $N_{l}$ are thrown out. We restrict the boundaries of the surface to
$V_{200} \ <$ 300 {\kms} to insure that gas in virialized halos has ample
time to cool and collapse to the disk (e.g. Rees \& Ostriker 1977)
by a redshift, $z$ = 2.6, the median
redshift of the kinematic sample. We also assume $dP_{CDM}$ cuts off below
{\vc}
= 30 {\kms}
since gas photoionized to  temperatures of
$\sim$ 10$^{4}$ K by the UV background radiation escapes from the
dark matter halos with {\vc} $<$ 30 {\kms} (Thoul \& Weinberg 1995; Navarro \& Steinmetz
 1997;
Kepner {\etal} 1997: we investigate the consequences of
modifying this restriction in $\S$ 6).
Figure~\ref{inputv200_bw}  shows the
resulting distributions of $V_{200}$
for all the CDM models in Table~\ref{tab:modparm}.
All the  CDM curves exhibit maxima
near the {\vc} cutoff predicted by hierarchical cosmologies,
and as predicted the curves decline with increasing {\vc}. As expected the largest
fraction of massive halos is indicated for the ICDM models. Thereafter,
the fraction
decreases progressively from OCDM, $\Lambda$CDM, to SCDM adiabatic
models.

\begin{table}[ht] \footnotesize
\begin{center}
\caption{Model Parameters} \label{tab:modparm}
\begin{tabular}{lccccc}
MODEL
& SCDM
& $\Lambda$CDM
& OCDM
& ICDM
&TF
\cr
\tableline
$\Omega_{M}$       & 1.0 & 0.3 & 0.2 &  0.2&0.3          \cr
$\Omega_{\Lambda}$ & 0.0 & 0.7 & 0.0 &  0.8&0.7          \cr
$h$                & 0.5 & 0.7 & 0.7 &  0.7&0.7          \cr
$\sigma_{8}$       & 0.6 & 1.0 & 1.03&  0.9& ...         \cr
$n$                & \tablenotemark{a} & \tablenotemark{a} & \tablenotemark{a} &$-$1.8&
...         \cr
\tableline
\end{tabular}
\end{center}
\tablenotetext{a}{CDM power spectrum given by Bardeen {\etal} (1986) expression}
\end{table}

\subsubsection{TF}

The curve labeled TF in Figure~\ref{inputv200_bw} represents the null hypothesis discussed
above. In this case the x-axis corresponds to  {\vm} rather than {\vc}.
This is because {\vc} is a theoretical construct, whereas the null hypothesis
is based only on the {\em observed} properties of current galaxies. The crucial relation
ship
here is
the Tully-Fisher equation which connects $L$ and {\vm}; i.e.,

\begin{equation}
L(V_{max}) = L_{*}(V_{max}/V_{*})^{\gamma}
\cmma
\end{equation}

\noindent where $V_{*}$ is {\vm} for a galaxy with luminosity $L$ = $L_{*}$.
With the last equation we can obtain expressions for
the interception probability

\begin{eqnarray}
dP_{TF}(z,V_{max}) =  \Bigl [ dz{\pi \over 2}(1+z)^3 
\bigl ( -c {dt \over dz} \bigr )
\Bigr ] \nonumber \\ 
\times \Bigl [ R_d^{Cor}(V_{max},z)N_0^{Cor}(V_{max},z) \Bigr ] ^2 
\nonumber \\
\times  dV_{max}n(V_{max},z)F(N_0^{Cor})
\cmma
\end{eqnarray}

\noindent where d{\vm}$n[L(V_{max})]$ is the present density of galaxies
with maximum rotation velocities in the interval  ({\vm}, {\vm}$+$d{\vm}),
and $F(N_{0}^{Cor})$ is given by eq. (11) with $N_{0}^{Cor}$ substituted for
$N_{0}^{iso}$.
To determine $n[L(V_{max})]$
we assume  a
Schecter luminosity function; i.e.,

\begin{equation}
dV_{max}n(V_{max},z) = {\Phi_{*}}({L / L_{*}})^{-\alpha}{\rm exp}(-L/L_{*})d(L/L_{*})
\perd
\end{equation}

\noindent The TF curve depends on the parameters $\alpha$, $\gamma$,
and $V_{*}$ for which we assumed values of 1.0, 3.0, and 250 {\kms}
respectively.
These are representative for the values of parameters
adopted by \cite{gonz00} who computed $n$({\vm},0). They analyzed
data sets from extensive surveys carried out at $B$ magnitudes.
However, using Cepheid calibrated galaxies \cite{sakai00} derive
Tully Fisher relations indicating lower values of $V_{*}$
($\approx$ 180 {\kms})
for the $B$ magnitude Tully Fisher relation. Low values of $V_{*}$, as well as
higher values of $\gamma$, are indicated by their $I$ band Tully Fisher relation,
and from similar relations found by \cite{giov97}.
In $\S$ 6 we discuss the sensitivity
of our results to these parameters . The point we wish to
emphasize is that
for acceptable ranges of these parameters the TF curve differs from
the CDM curves in that  it (a) peaks at {\vm} $\approx$ $V_{*}$/2 $\approx$ 130 {\kms},
which is large compared
to the 30 {\kms} peak of the CDM curves, (b) has little power near the latter
peak, and
(c) falls off exponentially when {\vc} $>$ $V_{*}$/2. Of course the comparison with CDM
is
inexact since $P_{CDM}$ depends on {\vc} while $P_{TF}$ depends on
{\vm}. Still the differences between
{\vc} and $V_{max}$ are not large enough to invalidate these conclusions.

\section{IONIZED GAS}

Assume that disks arise from the infall of ionized gas
predicted to fill dark-matter halos. According to the MM model
the halo gas is accreted during merger events with other halos
and
is shock-heated to the virial temperature of the halo,
$k{T_{vir}}$ $\equiv$ (1/2)(${\mu}m_{H}$$V_{200}^{2}$); we
assume $\mu$ = 0.4.
The duration of the
accretion phase is presumably short
compared to $t_{M}$, the time-interval
between events.
For the mass interval corresponding to $V_{200}$ = [30, 300] {\kms},
$T_{vir}$ ranges between
3$\times$10$^{4}$ and 3$\times$10$^{6}$K.
In the
case of massive halos the cooling time, $t_{cool}$, exceeds the
age of the gas, $t_{M}$,  at $r$ = $r_{200}$. Because the density of the
gas increases with decreasing radius,
$t_{cool}$ decreases with
decreasing radius until
$t_{cool}$ = $t_{M}$ at the cooling radius, $r_{cool}$.
At $r \ <$  $r_{cool}$  the gas moves radially
inward in a quasistatic cooling flow \citep{fabian94}.
Cool clouds form in pressure equilibrium with the hot
gas  since the hot gas is thermally unstable.
Due to the loss of buoyancy, the denser clouds fall inward at speeds
determined by the imbalance between gravity and ram pressure.
Because of rapid cooling, most of the gas in lower mass halos
cools before
moving inward, but within a limited range of
halo masses there is always hot low-density gas left over which also moves
inward in a cooling flow and exerts pressure on cooler clouds which have
formed. The MM hypothesis is that the cool clouds are photoionized by
background UV radiation and that they are the sites of C IV absorption
lines in QSO absorption systems. Here we extend this hypothesis
to model
the ionized gas causing C IV absorption
in damped {\lya} systems.

\begin{figure}[ht]
\includegraphics[height=3.8in, width=3.3in]{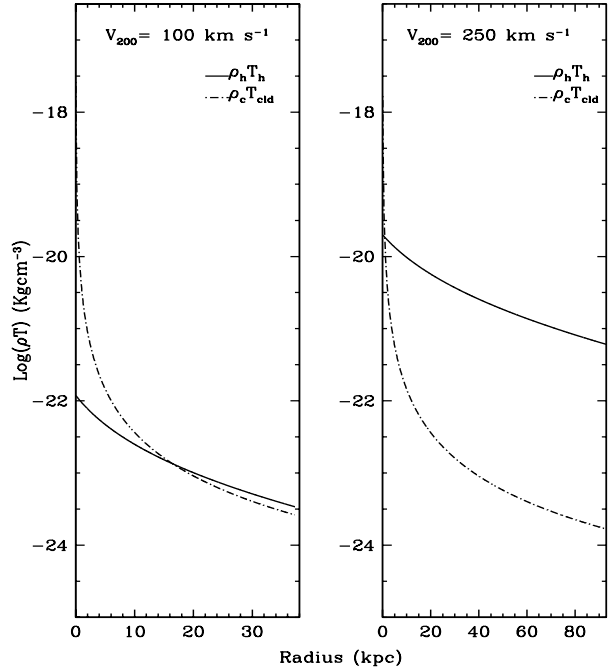}
\caption{Solid curves are pressure profiles for hot gas in halos with {\vc}
= 100 and 250 {\kms} in (a) and (b) respectively. Pressure
profiles are given by product of ${\rho(r)}_{h}$ and $T_{h}(r)$, the density and temperat
ure
of the hot gas.
Profiles computed according to prescription in Appendix B. Dot-dashed curves
are smooth density of cool gas times temperature of clouds; i.e. ${\rho}_{c}$$T_{cld}$.
When latter exceeds pressure of hot gas,
presure equilibrium between cool clouds and  hot gas breaks down (see $\S$ 4.1)}
\label{pressure_NFW}
\end{figure}

\subsection{Two Phase Structure of the Halo Gas}

Following MM we assume the hot gas is in hydrostatic equilibrium with the dark matter
potential and that it exhibits adiabatic
temperature and density profiles
out to {\rmin} = min({\rc},{\rv}).
At $r \ > r_{min}$ the hot gas either
follows isothermal profiles when {\rv} $>$ {\rc} or does not
exist when {\rv} $<$ {\rc}.
In Appendix B we derive expressions for {\rc}, and
for $\rho_{h}(r)$ and $T_{h}(r)$, the
density and temperature profiles for hot gas in halos corresponding to
the unmodified NFW rotation curves in eq. (7). Examples of
pressure profiles for high and low mass halos
are shown in Figure~\ref{pressure_NFW} . The Figure shows
the gas pressure in high mass halos to be
10 to 100 times higher than in low mass halos. The high pressures are
mainly
due to the higher virial temperatures of massive halos.

\begin{figure}[ht]
\includegraphics[height=3.8in, width=3.3in]{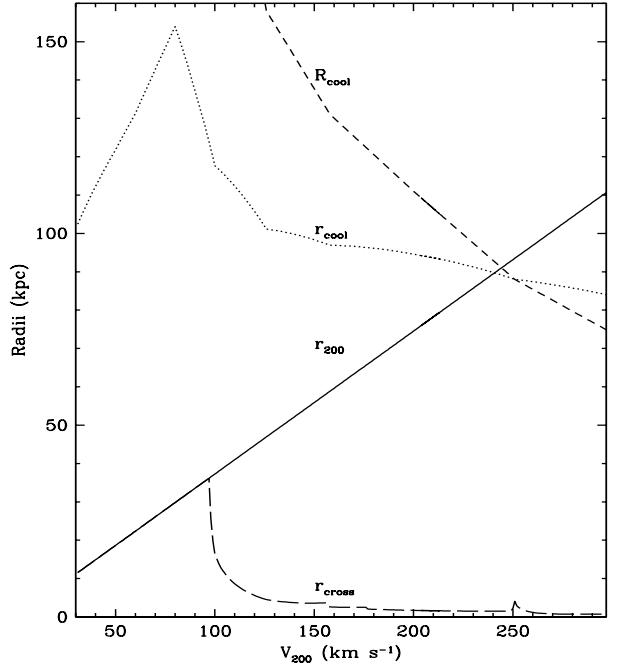}
\caption{Critical radii versus {\vc} for {$\Lambda$}CDM cosmology, NFW halos, and
$z$ = 2.5. $R_{cool}$, the cooling radius of an infinite isothermal sphere is
computed from eq. (B7), and $r_{cool}$ the cooling radius of an NFW halo is
computed from eq. (B6). The ''virial radius'' $r_{200}$ is computed from
eq. (3). The crossing radius $r_{cross}$, computed from eq. (C8),
is the radius within which pressure equilibrium breaks down. Note {\rmin}
= min({\rv},{\rc}) for values of {\vc} where {\rcr} $\le$ {\rv}. At
{\vc} $<$ 100 {\kms}, {\rmin} = {\rcr} = {\rv} (see $\S$ 4.2).}
\label{radii_NFW}
\end{figure}

Gas in
the cool phase comprises identical uniform spherical clouds with mass, $M_{cld}$,
radius, $R_{cld}$,  temperature,
$T_{cld}$, and internal density, $\rho_{int}$.  We adopt the MM model by
assuming clouds form
at $r=r_{min}$ with mass
$M_{cld}(r_{min})$ = 4$\times$10$^{5}$ $M_{\odot}$, and
temperature, $T_{cld}$ =2$\times$10$^{4}$K, since they
are assumed to be photoionized. To compute $R_{cld}$, MM assume
$\rho_{int}$ is set by pressure equilibrium with the surrounding hot gas; i.e.,
$\rho_{int}$ = ${\rho_{h}}T_{h}/T_{cld}$.
But this assumption is incorrect for a wide range of halo masses and radii.
Specifically, pressure equilibrium
breaks down when the pressure is sufficiently low for $R_{cld}$ to equal the
mean distance separating clouds. This will occur when
$\rho_{c}(r)$, the average density
of cool gas, exceeds $\rho_{int}(r)$; i.e., when ${\rho_{c}}{T_{cld}}$
$>$ ${\rho_{h}}{T_{h}}$.
According to MM,
$\rho_{c}(r)$ is given by

\begin{equation}
{\rho_{c}(r)} = {C_{1} \over r^{2}} \ ,
\end{equation}

\noindent where the constant $C_{1}$ is evaluated in Appendix C.
Figure~\ref{pressure_NFW}  demonstrates how $r_{cross}$, the radius at which
${\rho_{c}}{T_{cld}}$ = ${\rho_{h}}{T_{h}}$,
decreases with increasing {\vc}.
In Figure~\ref{radii_NFW}  we plot $r_{cross}$, {\rc}, and {\rv} versus
$V_{200}$ for a  ${\Lambda}$CDM cosmology, NFW
halos, and $z$ = 2.5 (we also plot the cooling radius of a singular isothermal
sphere, $R_{cool}$, for comparison with NFW halos).
The point of the Figure is to show that (1) pressure equilibrium
breaks down throughout halos with {\vc} $<$ 100 {\kms}, (2) pressure
equilibrium breaks down only near the centers of halos with {\vc}
$>$ 100 {\kms}, (3) {\rmin} = {\rv} for halos with {\vc} $<$ 250 {\kms},
and (4) {\rmin} = {\rc} for halos with {\vc} $>$ 250 {\kms}.
These trends  are qualitatively similar for all the
background cosmologies in Table~\ref{tab:modparm}.

In order to compute the structure of the cool clouds
we assume
${\rho_{int}}$ = ${\rho_{c}}$ at $r \ < r_{cross}$. As a result

\begin{equation}
\rho_{int}(r) = \left\{
\begin{array}{ll}
{{\rho_{h}(r)}T_{h}(r) \over T_{c}} \ ; r > r_{cross} \\
{\rho_{c}(r)} \ ; r < r_{cross}
\perd
\end{array}
\right.
\end{equation}

\noindent The cross-section of the infalling clouds is then given by

\begin{equation}
A(r) = {\pi}{\Biggl(}{3M_{cld}(r) \over 4{\pi}}{\Biggl)}^{2/3}\left\{
\begin{array}{ll}
({T_{c} \over {T_{h}(r){\rho_{h}(r)}}})^{2/3} \ : r > r_{cross} \\
({1 \over {\rho_{c}(r)}})^{2/3} \ ; r < r_{cross},
\end{array}
\right.
\end{equation}

\noindent which will be smaller than the value computed from pressure equilibrium
at $r \ < r_{cross}$.
At $r \ > r_{cross}$, $M_{cld}$ is assumed to be a function of $r$
as the clouds may sweep up hot gas as they fall inward (see
Benjamin \& Danly 1997 for a discussion of this problem). At $r \ < r_{cross}$
the density of hot gas is negligible, and so
we assume $M_{cld}(r)$ = $M_{cld}(r_{cross})$.
Therefore
we compute $M_{cld}(r)$
by solving

\begin{equation}
{dM_{cld} \over dr} = \left\{
\begin{array}{ll}
-A(r){\rho_{h}(r)} \ ; r > r_{cross} \\
0 \ ; r < r_{cross},
\end{array}
\right.
\end{equation}

\noindent  subject to the  boundary value of
$M_{cld}(r_{cross})$.
Explicit expressions for $A(r)$ and $M_{cld}(r)$ are given in
Appendix C.

\subsection{Cloud Kinematics}

The hot gas surrounding the clouds will exert a drag
force
opposing their radial infall.
Assuming the drag is caused
by momentum imparted to the clouds by hot gas
swept up during infall, we find the radial equation of motion to be

\begin{equation}
{d \over dt}{\Bigl[}M_{cld}V_{r}{\Bigr]}=M_{cld}g(r)
\cmma
\end{equation}

\noindent where $g(r)$ is the gravitational acceleration and the radial velocity, $V_{r}
= -dr/dt$,
is positive for infalling bodies. It  follows that

\begin{equation}
{dV^{2}_{r} \over dr}+{\Bigl(}{d{\rm ln}M_{cld}^{2} \over dr}{\Bigr)}V^{2}_{r}= -2g(r)
\cmma
\end{equation}

\noindent (which corresponds to the case $C_{D}$ = 2 of
Benjamin \& Danly [1997]). The solution to equation (21) is

\begin{equation}
V^{2}_{r}(r) ={2 \over M^{2}_{cld}(r)} {\int_{r}^{r_{start}}dr^{'}g(r^{'})M^{2}_{cld}(r^{
'})}+V_{r}^{2}(r_{start})
\cmma
\end{equation}

\noindent where we have computed the radial velocity of a cloud that starts
to infall from
an initial radius, $r$=$r_{start}$, with an initial velocity, $V_{r}(r_{start})$.
Note, this expression ignores fragmentation of the
clouds due to Kelvin-Helmholtz instabilities which
may be important \citep{lin00}.

To compute $V_{r}(r)$ we assume the clouds infall from rest
at $r$=$r_{min}$. As a result we solve eq. (22) with $r_{start}=r_{min}$
and $V_{r}(r_{start})$ = 0. The resultant $V_{r}(r)$ are valid in the
interval $r$=[$r_{cross}$,$r_{min}$]. The solution for $V_{r}(r_{cross})$
acts as boundary condition in the first of two scenarios we consider for
obtaining $V_{r}(r)$ at $r \ < r_{cross}$.

In the first scenario, clouds at $r$$<${\rcr} follow pressure-free ballistic trajectories
along which $M_{cld}(r)$=const. In halos with $r_{cross}$ less than $r_{min}$ we
solve eq. (22) by assuming $r_{start}=r_{cross}$ and let $V_{r}(r_{start})$
equal the $V_{r}(r_{cross})$ obtained from the solution at $r$ $>$ $r_{cross}$.
For halos with $r_{cross}$ greater than $r_{min}$
we solve eq. (22) by assuming $r_{start}=r_{min}$ and $V_{r}(r_{start})$=0; i.e.,
the clouds undergo ballistic infall from rest at $r$=$r_{min}$.

In the second scenario we assume the cloud kinematics at $r$ $<$ $r_{cross}$ are dominated
by random motions. These may be generated by feedback from supernova remnants arising from
star formation stimulated in cloud-cloud collisions. Such scenarios have been
suggested to solve the ``cooling catastrophe'' characterizing the hierarchical
build up of galaxy-scale structure in most CDM models \citep{white91}.
We obtain the velocity dispersion of the clouds by solving the Jeans equations
for ${\sigma_{r}}$, the radial velocity dispersion for a system of clouds with (i) an
isotropic velocity distribution, (ii) an average density distribution given by
$\rho_{c}(r)$, and (iii) the gravitational field of  an NFW halo (see eq. D4 in Appendix
D).
We then randomly
draw the velocities of individual clouds from  a Gaussian velocity distribution with dispersion
given by $\sigma_{r}$.

We emphasize that our model is most uncertain for
low mass halos. This is because
the underlying assumption of pressure equilibrium, which allowed MM
to compute the properties of the clouds,
breaks down {\em throughout} the infalling gas
for halos with {\vc} $<$ 100 {\kms}.
In these halos the properties of individual clouds are difficult to
compute because without a confining medium the clouds
become indistinguishable at $r$ $<$ $r_{cross}$.
Our approach to this problem is to
assume
the line of sight traverses a medium containing ``cloud-like''
structures
with fixed masses and that these give rise to C IV absorption lines.
This assumption needs to tested with
high-resolution hydrodynamical simulations of gas at $r$ $<$ $r_{cross}$
which is subject to input of mechanical energy.
We have also assumed eq. (17) is valid to obtain the average density of the cool
gas in every model. While this expression, which is based on
mass conservation for infalling gas, is physically justified  in the case
of systematic infall, it is arbitrary when the gas kinematics are dominated
by random motions.
Nevertheless we believe the results should provide
insights into the observational consequences of random motions (see
$\S$ 6).

\subsection{Absorption Properties of the Cool Gas}

In order to compute C IV absorption profiles produced by the
infalling clouds we select their locations along the line
of sight from
the cumulative interception probability function

\begin{equation}
F(<s)= {\int_{s_{min}}^{s}dy{\Biggl[}{{\rho_{c}(r(y))} \over M_{cld}(r(y))}{\Biggr]}A{\bigl(}
r(y){\bigr)} \over
  \int_{s_{min}}^{s_{max}}dy{\Biggl[}{{\rho_{c}(r(y))} \over M_{cld}(r(y))}{\Biggr]}A(r(y
))} \;\;\; ,
r={\sqrt {y^{2}+b^{2}}} \;\;\; ,
\end{equation}

\noindent  where the path integral propagates along a sightline with impact parameter
$b$ (where $b$ is the distance in the plane of the sky
separating the QSO sightline from the center of the galaxy), $s_{max} \ =
\sqrt{r_{min}^{2}-b^{2}}$, and $s_{min} = \ - s_{max}$.
We compute
$N_{CIV}^{cld}(r)$, the  C IV column density of a given cloud,
from the following expression:

\begin{equation}
N_{CIV}^{cld}(r)= X_{CIV}(r)N_{H}^{cld}
   \;\;\; ,
\end{equation}

\noindent where the total H column density, i.e. H$^{0}$ $+$ H$^{+}$, of the cloud is given
by

\begin{equation}
N_{H}^{cld}(r)= {3{\sqrt 2}M_{cld}(r)S \over 4{\mu}m_{H}A(r)}
   \;\;\; ,
\end{equation}

\noindent and $X_{CIV}(r)$, the ratio of
$C^{+3}$ to total hydrogen volume densities, is assumed to depend only on
$r$, the distance of the cloud from the center of the galaxy. To account
for variations in $N_{CIV}^{cld}$ caused by the variations of sightline
locations across the projected face of the cloud we have introduced the
uniform deviate $S$ which selects random numbers from the interval
$S$ =(0,$\sqrt{2}$).
We evaluate $X_{CIV}(r)$ by assuming

\begin{equation}
X_{CIV}(r)=X_{CIV}(R_{d})(R_{d}/r)^{\beta}
\perd
\end{equation}

\noindent To determine $X_{CIV}(R_{d})$ and $\beta$ we
compute the average C IV column density, ${\overline N_{CIV}(b)}$, where

\begin{eqnarray}
{\overline N_{CIV}(b)}= \int_{s_{min}}^{s_{max}}dsX_{CIV}(r(s)){\Biggl[}{{\rho_{c}(r(s))}
 \over {\mu}m_{H}}{\Biggr]}
   \;\;\; , {\nonumber} \\
r={\sqrt {s^{2}+b^{2}}} \;\;\; .
\end{eqnarray}

\noindent  We find that

\begin{equation}
{\overline N_{CIV}(b)}=K {\Biggl(}{R_{d} \over b}{\Biggr)}^{1+{\beta}}
\cmma
\end{equation}

\noindent where

\begin{equation}
K={2C_{1} \over {\mu}m_{H}}{X(R_{d}) \over R_{d}}{\int_{0}^{x_{max}}{dx \over (1+x^{2})^{
1+{\beta}/2}}}
\cmma
\end{equation}

\noindent and where $x_{max} = \ [(r_{min}/b)^{2}-1]^{1/2}$.
We then require the total  C IV column density in $n_{cld}$ clouds
to agree with ${\overline N_{CIV}(b)}$
within some accuracy, $\epsilon$; i.e., .

\begin{equation}
{\Biggl|}{{\sum_{i=1}^{n_{cld}}N_{CIV}^{cld}(r_{i})} \over {\overline N_{CIV}(b)}}-1{\Biggr|}
\ {\le} \ {\epsilon}
\cmma
\end{equation}

\noindent  where $\epsilon$ = 0.2 We adjust the input parameters $K$ and $\beta$ by comparing
empirical and model generated frequency distributions of C IV column densities
(see Figure~\ref{civcoldist}), and compute $X(R_{d})$ from eq. (29).

\section{RESULTS OF MONTECARLO SIMULATIONS: LOW-ION GAS}

\subsection{CDM}

Assume the low ion gas to be in disks with rotation
curves normalized to {\vc}
at $r$ = {\rv}, and
{\vc} to be randomly drawn from the distributions
in
Figure~\ref{inputv200_bw}.
In that case the kinematics of the low ion absorption profiles are
determined by the form of the rotation curve, {\vrot}, the value of the central
column density, $\Nperp$, and the thickness of the disks, {$h$}
(see PW1). Because the linewidth, {\dv}, increases with {$h$} we
follow PW1 by
adopting the largest plausible value, {$h$} = 0.3{\rd}, in order
to maximize {\dv}.
As stated in {$\S$} 2.3, {\vrot}
either equals {\vc} or {$f_{V}$}{\vc}.
We also make two assumptions about
$\Nperp$. Either $\Nperp$ equals  the central column density in
adiabatically contracted NFW halos, $N_{0}^{MMW}$,  or
$\Nperp = 10^{21.2}$~cm$^{-2}$
where the latter value
is meant to illustrate scenarios in which
star formation has
consumed most of the disk gas
(the value of $R_d$ does not affect
low ion kinematics since they are independent of
the absolute scale length [PW1]).
As a result the kinematics of the low ions are
represented by four independent
models.
Because each of these will be linked to  two kinematic models
of the ionized gas (discussed in $\S$ 5.2..1) we consider a total
of eight kinematic models for each cosmogony. Their properties
are summarized in Table~\ref{tab:models}.
The models are designated by 4 letters;
the first, $M$ or $N$, specifies whether $\Nperp = N_0^{MMW}$ or
10$^{21.2}$ cm$^{-2}$, the second, $V2$ or $fV$, specifies whether
{\vrot} = {\vc} or $f_{V}${\vc}, and the third, $R$ or $B$, indicates
whether the velocity field at $r$ $<$ $r_{cross}$ is dominated
by random motions or ballistic infall. Thus the model MV2B
has a disk with MMW central column density, disk rotation speed
given by {\vc}, and ballistic infall at $r$ $<$ {\rcr}.
Because the low ion kinematics of models MfVB, MV2B, NfVB, and
NV2B
are equivalent to those of models
MfVR, MV2R, NfVR, and NV2R
we shall
discuss low ion results only for the former group.

For a given {\vc} we find
our single-disk CDM  models with {\vrot} = $f_{V}${\vc} to result in
{\dv} that are larger than predicted by
Kauffmann (1996)
who did not correct for the gravitational contributions of the disk nor
for adiabatically
contracted halos to the disk rotation curves. Even so,
none of our CDM models
reproduces the observed low ion {\dv} distribution. This is contrary to the
expectations of MMW who conjectured that the higher {\vrot} produced by
the more realistic rotation curves would result in a {\dv} distribution
compatible with observation.
The reasons  this does not occur are
illustrated in
Figure~\ref{inputv200_bw}. In every case the median {\vc} of the
the intercepted disks is much less than 100 {\kms}.
Because the {\dv} predicted for rotating disks typically equals {\vc}/3,
the predicted median {\dv} will be much less than 50 {\kms}. By contrast
the median {\dv} of the observed distribution is about 80 {\kms}.

\begin{table*} \footnotesize
\begin{center}
\caption{Disk-Halo Models} \label{tab:models}
\begin{tabular}{lccccccccc}
&\multicolumn{4}{c}{$N_0$ = $N_0^{MMW}$ or $N_0^{Cor}$ 
\tablenotemark{a}}& \multicolumn{4}{c}{$N_{0}$=10$^{21.2}$cm$^{-2}$}& \\
\cline{2-5} \cline{7-10}
&\multicolumn{2}{c}{$V_{rot}$=$V_{200}$}& \multicolumn{2}{c}{$V_{rot}$=$f_{V}$$V_{200}$}
&& \multicolumn{2}{c}
{$V_{rot}$=$V_{200}$}& \multicolumn{2}{c}{$V_{rot}$=$f_{V}$$V_{200}$}  \\
\cline{2-5} \cline{7-10}
MODEL
& Ballistic
& Random
& Ballistic
& Random
&
& Ballistic
& Random
& Ballistic
& Random
\\
\cline{2-5} \cline{7-10}
MfVB&\nodata&\nodata&x&\nodata&&\nodata&\nodata&{\nodata}&{\nodata} \\
MV2B&x&\nodata&\nodata&\nodata&&\nodata&\nodata&\nodata&\nodata\\
MfVR&\nodata&\nodata&\nodata&x&&\nodata&\nodata&\nodata&\nodata\\
MV2R&\nodata&x&\nodata&\nodata&&\nodata&\nodata&\nodata&\nodata\\
NfVB&\nodata&\nodata&\nodata&\nodata&&\nodata&\nodata&x&\nodata\\
NV2B&\nodata&\nodata&\nodata&\nodata&&x&\nodata&\nodata&\nodata\\
NfVR&\nodata&\nodata&\nodata&\nodata&&\nodata&\nodata&\nodata&x\\
NV2R&\nodata&\nodata&\nodata&\nodata&&\nodata&x&\nodata&\nodata\\
\tableline
\end{tabular}
\end{center}
\tablenotetext{a}{Central column density 
perpendicular to the disk given by eqs.
(3) and (8) for CDM models ($N_{0}^{MMW}$) 
and by eq. (6) ($N_{0}^{Cor}$) for TF models}
\end{table*}

\begin{figure}
\includegraphics[height=4.8in, width=3.8in]{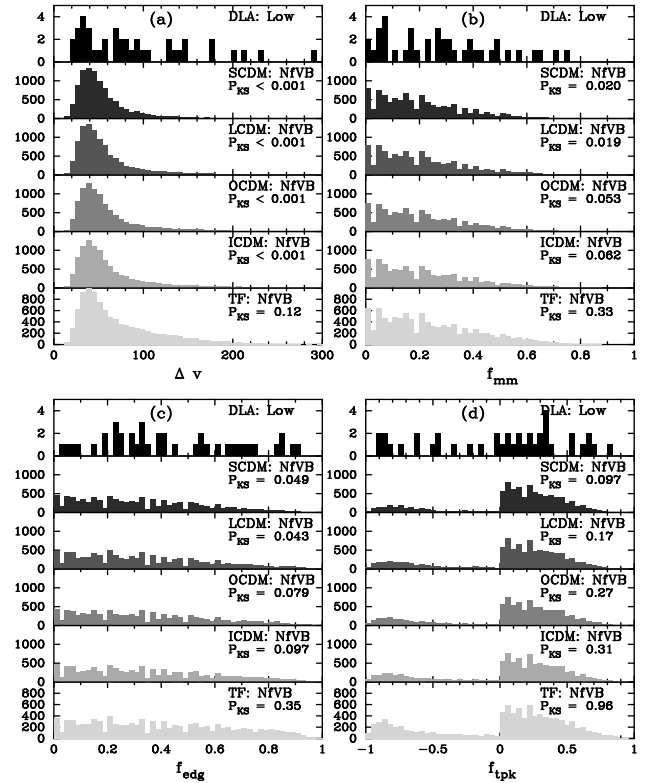}
\caption{Comparison between empirical and Monte Carlo distributions of low ion
test statistics for SCDM, {$\Lambda$}CDM, OCDM, ICDM, and TF cosmogonies. Results
for NfVB model compare distributions of (a) {\dv}, (b) $f_{MM}$, (c) $f_{edg}$, and (d) $f_{tpk}
$
test  statistics which are defined in paper I.
Briefly stated  these quantities are the (a) absorption velocity interval, (b)
the normalized difference between the mean and median velocity, (c) the difference between
the velocity of the strongest component and mean velocity, and (d) the difference between
the velocity of the second strongest component and mean velocity. The quantity
$P_{KS}$ is the KS probability that model and empirical distributions are drawn
from the same parent population.}
\label{Low_kinematics_E}
\end{figure}

Figure~\ref{Low_kinematics_E} shows the results
for model NfVB in each CDM cosmogony. With
{\vrot}=$f_{V}${\vc} and $\Nperp = 10^{21.2}$~cm$^{-2}$, this model
yields the best case results because it  generates the
largest {\dv} of the 4 models. The larger velocity widths follow from
(a) the higher rotation speeds, and (b) because the relatively low
central column density of model NfVB results in  smaller impact parameters out to the threshold
column density
{\nh}={\NL}, and smaller impact parameters cause larger {\dv} (PW1).
Nevertheless, application of the KS test shows that none of these
models is likely to fit the data (see Figure~\ref{Low_kinematics_E}).
This conclusion also holds for ICDM even though ICDM predicts a larger  fraction of
high {\vc}
halos than the other CDM models.
This is because
ICDM
is also an hierarchical model (Peebles 1999a, 1999b) and as a result a large fraction of
the protogalactic mass distribution is in  halos with
{\vc} $<$ 100 {\kms}.

\subsection{TF}

In the TF cosmogony we either let $\Nperp$ equal the empirically determined
$N_{0}^{Cor}$ or 10$^{21.2}$ cm$^{-2}$. As discussed previously
{\vrot} = {\vm} in all TF models, where
the distribution of {\vm} is shown in
Figure~\ref{inputv200_bw}.
Figure~\ref{Low_kinematics_E} also shows the model NfVB results
for TF.
Here the {\dv} test yields
$P_{KS}$({\dv}) = 0.12. While lower than $P_{KS}$({\dv}) = 0.65
exhibited by
models in which every halo has {\vrot} = 250 {\kms}
(PW1), this {KS} probability is sufficiently
large that the more realistic TF model cannot be excluded.
We shall check the robustness of these results for the TF and
CDM models in $\S$ 6.

\section{RESULTS of MONTECARLO SIMULATIONS: IONIZED GAS}

\subsection{Normalization}

In order to fit the simulations
to the data it is necessary to
specify the function
${\overline N_{CIV}(b)}$ where $b$ is the impact parameter. This function is crucial as it
constrains kinematic quantities
such as {\dciv} for the C IV profiles by fixing
the number
of clouds per line of sight (see eq. 30).
We normalize
${\overline N_{CIV}(b)}$  by determining
the parameters  $K$ and $\beta$ (eqs. 28 and 29)
from comparisons between  model and empirically
determined frequency distributions of C IV column densities, $f(N_{CIV},z)$.
The latter is the product of the number of damped {\lya} systems per
unit ``absorption distance'', $dn/dX$, times $g(N_{CIV})$, the conditional
distribution of  C IV column densities
given the presence of a
damped {\lya} system. We find that

\begin{eqnarray}
f(N_{CIV},z) = {dn \over dz}{\Biggl(}{dX \over dz}{\Biggr)}^{-1}g(N_{CIV}),
{\nonumber} \\
\ {dX \over dz} = {{(1+z)^{2}} \over {[(1+z)^{2}(1+{\Omega_{M}}z)-z(z+2){\Omega_{\Lambda}}]^{1/2
}}}
\perd
\end{eqnarray}

\noindent Because damped {\lya} systems are H I selected, the function $g$  will depend on
the differential area of the inclined H I disks giving rise to damped
{\lya} absorption. As a result the $g$ will depend on impact parameter,
$b$, and hence on $K$ and $\beta$ through eqs. (28) and (29).
We adjust $K$ and $\beta$
by comparing model and empirical $f(N_{CIV},z)$'s.

\begin{figure}[ht]
\includegraphics[height=3.8in, width=3.3in]{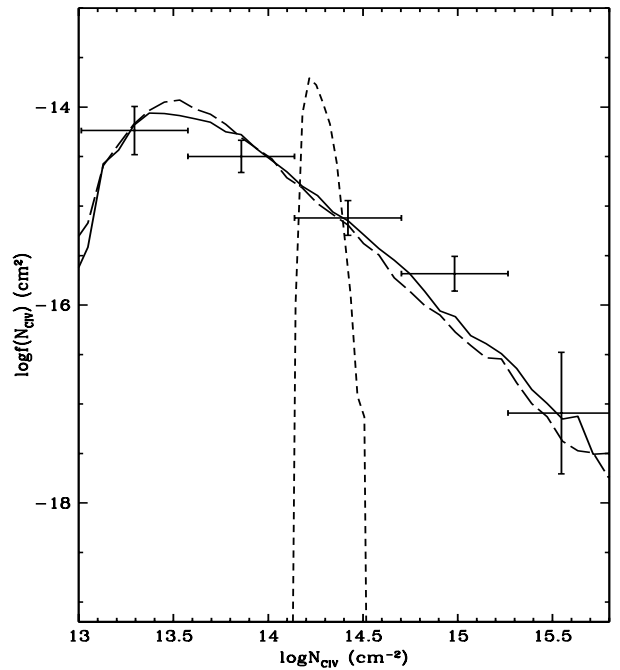}
\caption{Comparison between empirical and model CIV column density distributions.
The points with error bars are data. The solid and short dashed curves are
predictions for MfVB and NfVB models, $\beta$ = 1.5 and the {$\Lambda$}CDM cosmogony. The
short dashed curve corresponds to the MM model for halos
with {\vc} $>$ 150 {\kms} and in which $\beta$ $\approx \ -1$}
\label{civcoldist}
\end{figure}

We constructed the empirical $g(N_{CIV})$ from the 32 C IV column
densities inferred from the
profiles in Figure 1 in Paper I (The actual
column densities are reported in Prochaska \& Wolfe 1999 and in Lu {\etal} 1996).
To obtain the empirical $f(N_{CIV},z)$
we adopted the $\Lambda$CDM model and let $dn/dz$ = 0.22 which is appropriate
for the mean redshift of this sample. The results are shown as
points with error bars in
Figure~\ref{civcoldist}. We  compare this with the predictions for
models MfVB  and NfVB in the case of a $\Lambda$CDM cosmology.
The results are valid for all MXXX and NXXX models respectively. This is because
$f(N_{CIV})$ depends on the distribution of impact parameters,
but is independent of rotation speed and infall kinematics.
In both classes of models,  $\beta$ = 1.5 provides a good fit to the data, while
$K$ = 2$\times$10$^{15}$ cm$^{-2}$  for model  MfVB and 2$\times$10$^{14}$ cm$^{-2}$
for model NfVB.
When
MM computed ${\overline N_{CIV}(b)}$
for cool halo gas
photoionized by background radiation, they found
K and $\beta$ to vary with {\vc}, in contrast to our assumption of
uniform $K$ and $\beta$ for halos of all mass.
Fortunately, the C IV velocity profiles are independent of $K$
because both ${\overline N_{CIV}}$ and $N_{CIV}^{cld}$ are linearly
dependent on $K$, and as a result $K$ drops out of the determination
of $n_{cld}$ which is crucial in determining the velocity profile widths
(see eq. 30). On the
other hand the profiles do
depend on $\beta$.  Figure~\ref{civcoldist}
shows that $\beta$ = $-$ 0.9, which corresponds to
the MM results for halos with {\vc} $>$ 150 {\kms}, results
in poor fits to the data.  Therefore, the
$\beta$ $\approx$ 1.5 assumption is consistent with the
C IV data and is used in the calculations which follow.

\subsection{ Test Statistics}

\subsubsection{ CDM Models}

\begin{figure}[ht]
\includegraphics[height=3.8in, width=3.3in]{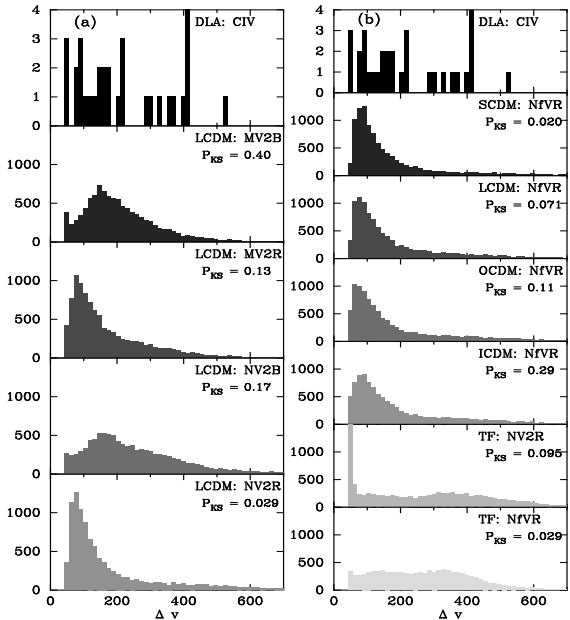}
\caption{Comparison between empirical and Monte Carlo distribution of
{\dciv}. (a) Results for {\lcdm} cosmogony and MV2B, MV2R, NV2B, and NV2R models.
(b) Results for SCDM through ICDM cosmogonies and NfVR model, and for TF cosmogony
with NV2R and NfVR models}
\label{civdelvtest}
\end{figure}

In Figure~\ref{civdelvtest}a
we compare the empirical {\dciv} distribution with
predictions by the $\Lambda$CDM cosmogony. We only show
results for the case {\vrot}={\vc}, since halo kinematics should
be unaffected by disk rotation speed for halos of a given mass.
We ignore results for the
$f_{mm}$, $f_{edg}$, and $f_{tpk}$  statistics
because in the case of C IV kinematics {\dciv} is the most
sensitive test statistic for testing
any of the models.
The principal difference between the distributions in Figure~\ref{civdelvtest}a
stems from the different velocity fields at $r$ $\le$ $r_{cross}$. Models with
ballistic infall (MV2B and NV2B) predict larger median {\dciv} than models with random
motions (MV2R and NV2R). For a given $\Nperp$ the ballistic models are in slightly better
agreement with the data because the larger {\dciv}'s are closer to
the observed values.

The value of $\Nperp$ also causes differences between the distributions. Comparison
between ballistic infall models MV2B and NV2B shows that MV2B predicts lower median {\dciv}
than NV2B. This is because the higher $\Nperp$ of model MV2B results in larger impact parameters
then cause the sightlines to sample the halos at larger radii where the infall
velocities are smaller. In this case the lower {\dciv} of the
MV2B model is in better agreement with the data.
At the same time model MV2R
is in better agreement with the data than NV2R, because the smaller
impact parameters predicted by the latter model result in more sightlines
traversing the
$r$ $\le$ $r_{cross}$ region where random motions give
rise to {\dciv} that are also lower.

The third difference between the {\dv} distributions depends on the assumed cosmogony.
In Figure~\ref{civdelvtest}b we use the NfVR model to illustrate the effects
of the assumed cosmology.
The cosmogonies differ according to the fraction of halos with large
{\vc}, a fraction which increases along the SCDM $\rightarrow$ ICDM sequence.
In every cosmogony in this Figure
a ``spike'' in the {\dciv} distribution at {\dciv} = 50 {\kms}
is present and increases in strength along the  sequence.
The spike arises
from sightlines traversing the outer regions of high-mass halos at large impact
parameters. The high pressure of hot gas in massive halos (see  Figure~\ref{pressure_NFW})
compresses the clouds thereby reducing their cross-sections. For clouds
of fixed mass the result is an increase
in column density. In most cases only one cloud is required to satisfy eq. (30)
at large $b$ where ${\overline N_{CIV}}$ is small. The spike occurs at
{\dciv} = 50 {\kms} because this is the FWHM of the profile caused by a single cloud with
an assumed internal velocity dispersion, $\sigma_{int}$ = 25 {\kms}. In fact
this value of $\sigma_{int}$ was chosen to reproduce the narrowest C IV profiles in
our sample.

It is worth noting that despite their differences most of the models yield large
$P_{KS}$({\dciv}) values. This is in contrast to tests of low ion kinematics. In that case
all of the $P_{KS}$({\dlow}) values were too small; i.e.,  none of the models was compatible
with the
observed distribution of low ion {\dv}. This could imply
that while the infall interpretation for the high ions is correct,
the disk interpretation for the
low ions is incorrect.

\subsubsection{ TF Models}

Figure~\ref{civdelvtest} b also compares the data with the predictions of the TF cosmogony.
In this case {\vc} equals  {\vm} or {\vm}/$f_{V}$ where the {\vm} are the input
disk rotation speeds.
As a result the CDM degeneracy of {\dciv} with respect
to disk rotation speed is broken in the TF models.
We show results for some examples to illustrate this effect.
As expected the {\dv} = 50 {\kms} spike is highest for the NV2R model
where {\vc} = $V_{max}$. Because of the larger fraction of halos with high
{\vc}, the median {\dv}'s are higher than in the CDM models.
Agreement with the empirical {\dciv} distribution
improves
in the case of random motions
at $r \ < r_{cross}$ and when {\vc} = {\vm}/$f_{V}$; i.e.,
with the NfVR model (note the higher P$_{KS}$({\dciv}) value
of the NV2R model is an artifact due to the large amplitude spike
at {\dciv} = 50 {\kms}), but the $P_{KS}$({\dciv}) values are
still leass than 0.05.

\subsection{Correlation Tests}

\begin{figure}[ht]
\includegraphics[height=3.8in, width=3.3in]{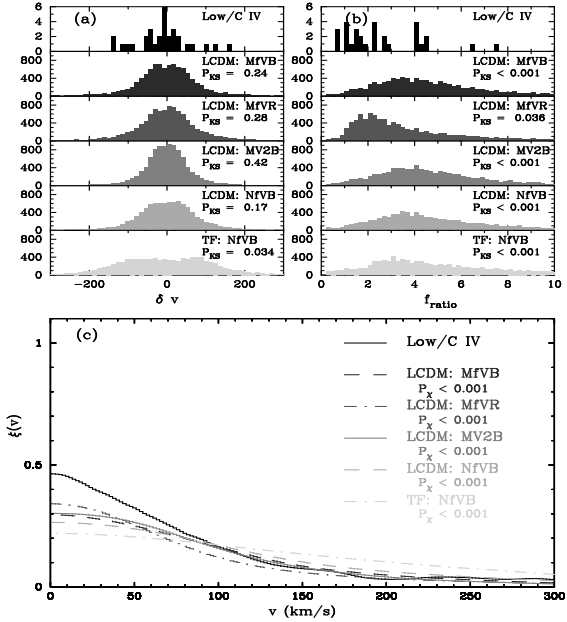}
\caption{Correlation Tests. (a) Comparison between empirical and Monte Carlo
distributions of {\dvm} for {\lcdm} cosmogony and MfVB, MfVR, MV2B,and  NfVB models.
The NfVB model is used with the TF cosmogony. (b) Same as (a) but $f_{ratio}$ is substituted
for {\dvm} statistic. (c) Comparison between empirical and Monte Carlo predictions
of $\xi(v)$ for C IV versus Low. Same models as in (a) and (b) are used. Curves are
data (dark solid curve), {\lcdm} MfVB model (dark dashed curve), {\lcdm} MfVR model (dark
dot-dashed curve, {\lcdm} MV2B model (light solid curve),  {\lcdm} NfVB model
(light dashed curve), and TF NfVB model (light dot-dashed curve)}
\label{corrmodeltest}
\end{figure}

In Figure~\ref{corrmodeltest} we compare empirical and predicted
distributions of {\dvm} and $f_{ratio}$. These  are the differences between the mean
velocities of the
C IV
and low ion velocity profiles   and  the ratio {\dciv}/{\dlow},
respectively . Comparison between empirical and predicted cross-correlation functions for
the C IV versus low ion velocity profiles is also shown.

$\bullet$ {\em {\dvm} Test:}

In CDM
all the models pass the {\dvm} test at more than 83$\%$ confidence.
In Figure~\ref{corrmodeltest}a we use $\Lambda$CDM to
illustrate the effects of (a) disk rotation speed, (b) impact
parameter, and (c) halo velocity field. Comparison between the MfVB and
MV2B models
shows that disk rotation speed is the most important effect. Specifically
$P_{KS}$({\dvm}) increases significantly when {\vc} decreases from
$f_{V}${\vc} to {\vc}. This behavior is straightforward to explain.
In the disk-halo models
the   half-width of the {\dvm} distribution  is roughly equal to
the sample average
of {\vrot}sin($i$). This is because the prototypical C IV profile
comprises two widely separated absorption components symmetrically displaced about
the systemic velocity of the halo. As a result the mean velocity of the
C IV profile equals
the systemic velocity. On the other hand
the low ion profile consists of  multiple contiguous components comprising a single feature
that
is displaced to {\em either} side of the systemic velocity.  In this case
the mean velocity of the low ion profile
is separated by $\sim$ {\vrot}sin($i$) from the systemic velocity.
Consequently the $\approx$ 75 {\kms}
half-width of the observed distribution limits
the fraction of rapidly rotating disks. This constraint
is especially severe for the TF models where
rotation speeds exceeding 200 {\kms} are typical.

On the other hand the effects  of impact parameter are not as significant. This is
because impact parameter affects the width of the velocity features rather
than the location of their velocity centroids. This explains why the
the MfVB and NfVB results are so similar. Furthermore, the effect of halo
velocity field is even less important as shown by comparison between the MfVB and MfVR
results. This tells us that for a sufficient number of C IV clouds,  the location of the
velocity  centroid of the C IV profile is independent of whether the clouds are
infalling or moving randomly. Therefore, {\dvm} is set by the magnitude
of {\vrot}.

$\bullet$ {\em $f_{ratio}$ Test:}

A natural consequence
of the disk-halo hypothesis is
the
prediction $f_{ratio}$ $\ge$ 1.
Because the infall velocities
of the high ions and the rotation speeds of the low ions occur
in the same potential well, they both scale linearly with {\vc}.
However, owing to projection effects, the line-of-sight velocity
gradients due to radial infall will exceed those due
rotation.
Thus, {\dciv}
will be larger than {\dlow}.
But the ratios are too large in
most CDM models because of the small {\dlow}
and the larger {\dciv}.
As a result
the best case models are those with large {\vrot} for the disks  and random motions
at $r$ $\le$ $r_{cross}$ for the halos. The best
model is  MfVR, as shown
in Figure~\ref{corrmodeltest} b.
The TF models also produce $f_{ratio}$ that are too high
because in most halos the sightlines intercept the region $r \ > r_{cross}$
where large infall velocities are present.

$\bullet$ {\em Cross-Correlation Function:}

None of the models, neither CDM nor TF, predicts a cross-correlation function with large
enough amplitude to fit the data.
The reason is insufficient overlap in velocity space between
the C IV and low ion profiles. In
Figure~\ref{corrmodeltest} c the
best case model is MfVR, indicating more overlap
occurs when the C IV clouds experience
random motions at $r$ $\le$ $r_{cross}$.
Even better agreement is obtained with model MV2R (not shown)
implying that overlap increases when the rotation
speed of the disk is reduced.
This interpretation is supported by
the results for the TF models which exhibit the worst agreement with the
data. The TF models predict
the largest fraction of halos with high
{\vrot} and lowest fraction of low-mass halos in which random motions dominate the
the velocity field
at $r$ $<$ {\rcr}.

\subsection{Model Summary}

In summary we find the following results for the models we haved tested so far:

$\bullet$
While the extent of the low ion {\dv} distribution rules out the single-disk
semi-analytic CDM models, it is
compatible with the TF models (see also Jedamzik \& Prochaska 1998).

$\bullet$
The C IV {\dv} distribution is compatible with most of the CDM models. In
the example shown in Figure 7 the best
agreement with the data occurs for models with (a) ballistic infall at $r$ $\le$
{\rcr} and (b) central disk column densities given by $N_{0}^{MMW}$. The best agreement
with the TF models is for  ballistic infall, $N_{0}$=$N_{0}^{MMW}$, and {\vc}
= {\vm}/$f_{V}$.

$\bullet$
The {\dvm} distribution is compatible with all of the CDM models, but is too narrow for the
TF models. This is a reflection of the low {\vrot} predicted by CDM and the high {\vrot} 
predicted
by the TF models.

$\bullet$
Most of the CDM  models predict $f_{ratio}$ distributions with median
values that are too large. This stems from the low values predicted
for {\vrot}. The same problem holds for the TF models, but
in this case the high $f_{ratio}$ stems from the
high values of {\dciv}. In both cosmogonies some models cannot be ruled out.

$\bullet$
None of the models predicts C IV versus low ion cross-correlation functions
in agreement with the data.

\section{PARAMETER TESTS}

The results of the last section may be summarized as follows: When the free parameters
of a given model are adjusted to satisfy one test, the model inevitably fails
a different test. If this is a generic feature of the disk-halo models, then they may
not apply to the damped {\lya} systems. This is an important conclusion
and we wish to determine how robust it actually is.
Indeed the models are characterized by several
free parameters that are not well determined, and it is possible we have not
found the optimal set. For this reason we now investigate the sensitivity
of our conclusions to variations of these parameters.

In a series of trial runs we found the model kinematics
to be most sensitive to three
parameters. The first is
the central perpendicular column density, $\Nperp$. Because the range of
impact parameters is limited by $\Nperp$, 
it influences both {\dciv} and {\dlow}.
To test the dependence of the kinematics on this quantity we simulate
H I disks with $\Nperp$
ranging between 10$^{20.8}$ and 10$^{22.2}$ cm$^{-2}$. These values
cover the low column densities of the $NXXX$ models
and approach the high values of the $MXXX$  models. The second parameter
is {\vcut}, the low-end cutoff  to the distribution of input {\vc} in the CDM models
or of input {\vm} in the TF models.
The kinematic results should be sensitive to {\vcut}
especially in the case of CDM  where {\vcut} is at
the peak of the  {\vc} distribution
(see Figure ~\ref{inputv200_bw}). The value used in our models,
{\vcut} = 30 {\kms}, is imposed by thermal expansion of photoionized gas
out of the potential wells of halos with
{\vc} lower than this (Thoul \& Weinberg 1995).
However
feedback due to supernova explosions
might drive gas out of disks with {\vc} as large as  100 {\kms} (see Dekel \& Silk 1986,
but see MacLow \&
Ferrara 1999).
For these reasons we let {\vcut} vary between 30 and 120 {\kms}.
The third parameter is the C IV column
density per cloud, $N^{cld}_{CIV}$.  The definition given in eqs. (25)$-$(27)
is for a  spherical cloud of a given mass and C$^{+3}$/H
ratio, i.e., $X_{CIV}$.
To account for
deviations from spherical symmetry or from our definition of
$X_{CIV}$ (see eq. 26)
we introduce the parameter $q_{CIV}$ which
is the ratio of the true C IV column density to our model definition.
We let $q_{CIV}$ vary between 0.1 and 1.3.
In order to supply the total C IV column density required at a given impact parameter,
the number of clouds must increase as $q_{CIV}$ decreases.
This  affects the  C IV kinematics
because {\dciv} will increase with cloud number.

\subsection{CDM}

\begin{figure}[ht]
\includegraphics[height=3.8in, width=3.3in]{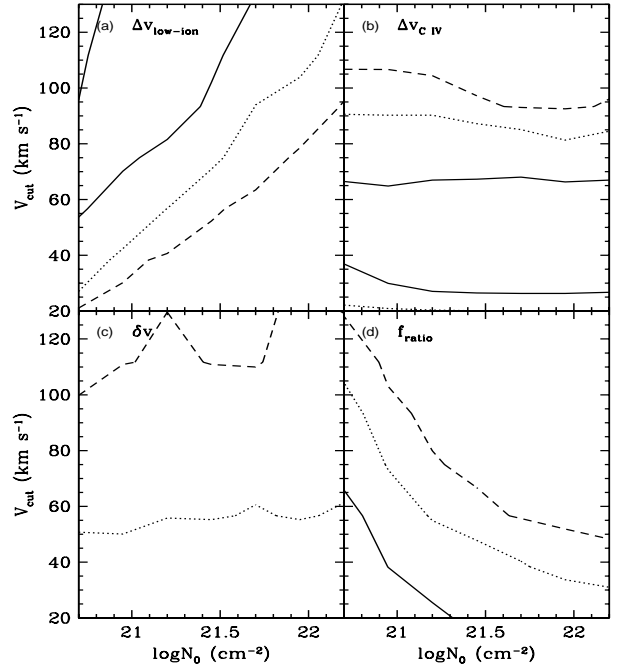}
\caption{Iso-probability contours in {\vcut} versus $\Nperp$ plane
resulting from parameter search of {\lcdm} cosmogony
where  NfVR model is assumed
in which  is a free parameter and where $q_{CIV}$ = 1.0.
Contours correspond to
$P_{KS}$ =  0.32 (solid lines), 0.05 (dots), and 0.01 (dashes).
Tests are (a) {\dlow}, (b) {\dciv}, (c) {\dvm}, and (d) $f_{ratio}$.}
\label{confidencelevels}
\end{figure}

The results of the parameter tests are summarized in
Figure ~\ref{confidencelevels} which shows iso-probability
contours
in the {\vcut} versus $\Nperp$ plane.  The
contours correspond to 0.01, 0.05, and 0.32
for $P_{KS}$({\dv}$_{low}$) (9a),
$P_{KS}$({\dv}$_{C IV}$) (9b),
$P_{KS}$({\dvm}) (9c), and $P_{KS}$($f_{ratio}$) (9d).
We show results for $q_{CIV}$ = 1.0 since smaller values
are found to
result in {\frat}
that are too large.
We choose a variant of the NfVR model in which
$\Nperp$ is a free parameter. We abandon ballistic infall
in favor of random motions at $r$ $<$ {\rcr} because we find
that random motions
produce more overlap in velocity space between low ion and  C IV
profiles and as a result
they produce better agreement
between model predictions and the empirical test statistics.

Figure ~\ref{confidencelevels}a shows the results for
the low ion {\dv} test. As expected models with the standard value {\vcut} = 30 {\kms}
are improbable for reasonable values of $\Nperp$.
Rather {\vcut} increases  along the $P_{KS}$({\dlow}) = 0.05
contour
from 35 {\kms} at $\log \Nperp = 20.8$~cm$^{-2}$ to 105 {\kms}
at $\log \Nperp = 22.0$~cm$^{-2}$.
An increase in
$\Nperp$ means larger impact parameters which in turn
imply smaller {\dlow} (see $\S$ 5.1).  Therefore, an increase in {\vcut} must
accompany the increase in
$\Nperp$ to boost the fraction of high-{\vc} halos required
to maintain the extent of the {\dlow} distribution.
By contrast
Figure ~\ref{confidencelevels}b shows that {\vcut} hardly varies
with  $\Nperp$ along the $P_{KS}$({\dv}$_{CIV}$) contours.
In our CDM models
most sightlines traverse halos with low {\vc} where cloud motions
in  NfVR models are dominated by random velocities drawn from
a Gaussian with
dispersion that is an insensitive function of radius (eq. D4).
Consequently {\dciv} will be independent of impact parameter
and hence independent of $\Nperp$. Therefore,
{\vcut} need not vary with $\Nperp$ to maintain the extent of the
{\dciv} distribution.

Figure ~\ref{confidencelevels}c and d  shows the iso-probability contours for the
{\dvm} and
$f_{ratio}$ tests. In Figure ~\ref{confidencelevels}c
none of the contours
rises
to $P_{KS}$({\dvm}) = 0.32  in the {\vcut} versus $\Nperp$ plane.
The shape of the $P_{KS}$({\dvm}) = 0.05 contour has the following implications.
First, the insensitivity of the contour to $\Nperp$ 
just indicates that the displacement
of the low ion velocity centroid from the systemic velocity of the
galaxy is determined by rotation speed
rather than {\dlow}. Second,
models with
{\vcut} $>$ 55 {\kms} are highly unlikely as they produce
{\dvm} that are too large.
In Figure ~\ref{confidencelevels}d,
$\Nperp$ decreases with increasing
{\vcut} along all iso-probability contours.
Because {\dciv} increases with increasing {\vcut},
$\Nperp$ must show
a corresponding decrease
to boost {\dlow}.
Otherwise $f_{ratio}$ becomes larger than observed.

\begin{figure}[ht]
\includegraphics[height=3.8in, width=3.3in]{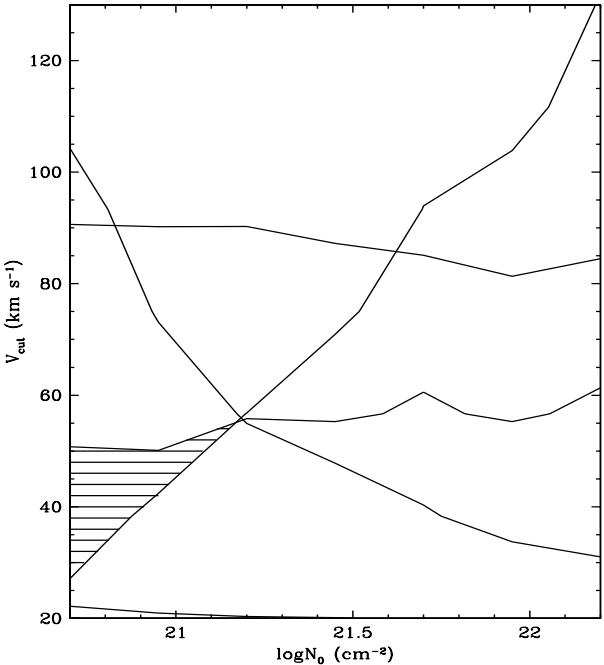}
\caption{solid lines are contours corresponding to $P_{KS}$ = 0.05
in Figure 9. The region in which $P_{KS}$ $\ge$ 0.095
for all 4 tests is denoted by horizontal lines}
\label{confidencelevels95}
\end{figure}

\begin{figure}[ht]
\includegraphics[height=3.8in, width=3.3in]{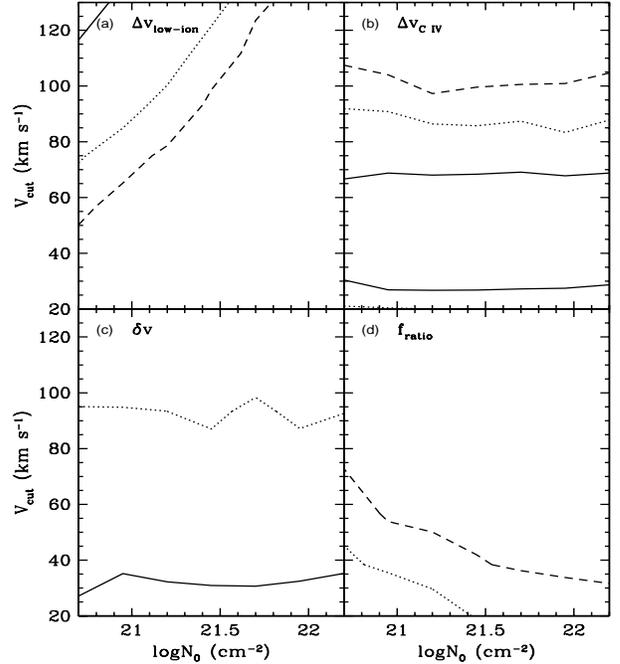}
\caption{Same as Figure 9, except with NV2R model}
\label{confidencelevelsV2R}
\end{figure}

Figure ~\ref{confidencelevels95} shows the $P_{KS}$ = 0.05 contours
from the previous Figure. The horizontal
lines trace out the region in the {\vcut} versus $\Nperp$
plane in which $P_{KS}$ $\ge$ 0.05 for all 4 tests; i.e.,
the parameter space of acceptable models.
Physically,
the resulting  range of {\vcut} (i.e. 35 to 50 {\kms}) is
acceptable for models in which gas photoionized by ionizing
background radiation escapes from low-mass halos (Thoul \& Weinberg
1995; Navarro \& Steinmetz 1997). On the other hand the upper limit on
$\Nperp$ (i.e., $\log \Nperp < 21.2$~cm$^{-2}$) 
may be too low to explain the shape of the
H I column-density distribution function (see $\S$ 7).
Furthermore, this Figure indicates these models may not be viable
as they occupy a small fraction
of the depicted parameter space.  Notice that the horizontal {\dvm} contour
at {\vcut} $\approx$ 50 {\kms}
is crucial in restricting the acceptable region to such a small area.
Because {\dvm} is set by rotation speed we investigated the
NV2R models to determine whether the lower rotation speeds would
enlarge the acceptable region. The results are shown in
Figure ~\ref{confidencelevelsV2R}. As predicted the lower
rotation speeds lift the restricting {\dvm} contour from
$\approx$ 50 to $\approx$ 95 {\kms}. However, the lower
rotation speeds also increase $f_{ratio}$, with a consequent
lowering of the $P_{KS}(f_{ratio})$ contour in  Figure ~\ref{confidencelevelsV2R}d.
Consequently there is no region in the {\vcut} verus $\Nperp$
plane in which all 4 tests result in $P_{KS}$ $\ge$ 0.05 for the
NV2R model. Therefore,
our conclusion concerning the size of acceptable regions in parameter space
appears to be robust.

Turning to the cross-correlation function we find
that none of the models within the
range of {\vcut} and
$\Nperp$ depicted in
Figure ~\ref{confidencelevels} results in
C IV versus low ion
cross-correlation functions with acceptable {$\chi^{2}$} values.
Apparently the combination of radial infall and disk rotation
produce C IV and low ion absorption profiles with insufficient overlap in
velocity space to explain the data. Better agreement is obtained when
$\log \Nperp > 23$~cm$^{-2}$. But column densities this high
are ruled out by the other tests. We shall return to this
dilemma in $\S$ 7.

In summary, by varying {\vcut}, $\Nperp$, and $q_{CIV}$  we find
regions of parameter space where
the {$\Lambda$}CDM models are in better agreement with the data than
for the ``standard'' values of the parameters adopted above. This
is especially true for the low ion {\dv} test which ruled out most
of the ``standard'' models (where {\vcut} = 30 {\kms}). However,
the models may still not be viable because of the restricted range
of allowable parameters, and because
none of the models is compatible with the C IV versus low ion cross-correlation
function.

\subsection{TF}

\begin{figure}[ht]
\includegraphics[height=3.8in, width=3.3in]{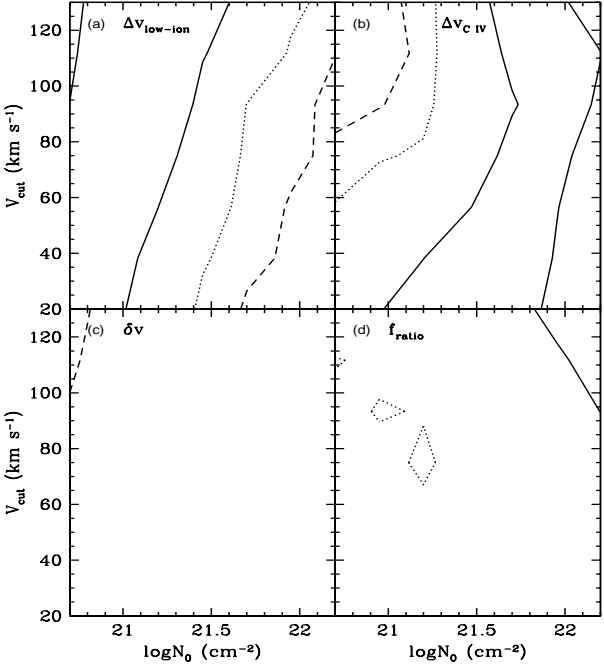}
\caption{Same as Figure 11, but for TF cosmogony}
\label{confidencelevels_TF}
\end{figure}

The corresponding results for the TF models are shown in
Figure ~\ref{confidencelevels_TF}. In this case the best fit
value of $q_{CIV}$ equals 1.3.
Figure ~\ref{confidencelevels_TF}a
illustrates the results for the low ion
{\dv} test. The Figure shows
the $P_{KS}$({\dv}$_{low}$) = 0.05 contour to enclose a larger
area of parameter space than in the CDM case. As in CDM,
{\vcut} increases with $\Nperp$ along iso-probability contours.
However,
the {\dlow} contours are less sensitive to {\vcut} because
the input halo distribution does not  peak at {\vcut} as in
CDM (see Figure 2).
Figure ~\ref{confidencelevels_TF}b shows that
by contrast with CDM the $P_{KS}$({\dv}$_{CIV}$) contours
are sensitive functions of $\Nperp$. This is because in the TF model
more sightlines traverse high {\vc} halos where C IV clouds undergo infall
at $r$ $>$ {\rcr},
and as a result {\dciv} is a sensitive function of impact parameter, and therefore of $\Nperp$.

The results for the {\dvm} and $f_{ratio}$ tests are plotted
in Figure ~\ref{confidencelevels_TF}c and d.
All
of the models result in $P_{KS}$({\dvm}) $<$
0.05. 
Clearly the large fraction of rapidly rotating disks  encountered in the TF model
produces {\dvm} that are too high. By contrast the results for
the $f_{ratio}$ lead to
$P_{KS}$({\dvm}) $>$ 0.05 throughout the parameter space
depicted in the Figure except for the upper right portion
where  $P_{KS}$({\dvm}) $>$ 0.32.
But part of the improvement
here is caused by spikes in the  {\dciv} distribution near
50 {\kms} which prevent $f_{ratio}$ from exceeding
observed values when the {\dlow} are small. While the physical basis for such spikes
is understood (see $\S$ 5.2.1), they have not been confirmed
by the data. As in CDM, none of the TF models produce C IV versus
low ion cross-correlation functions with acceptable $\chi^{2}$ values.
In fact the higher fraction of disks with large rotation speeds predicted by the TF
model increases the displacement between low ion and C IV
profiles which produces even lower cross-correlation
amplitudes than in the CDM models.

We also tested the sensitivity of the TF model results  to variations
of  (a) $\gamma$ and $V_{*}$, the power-law exponent and fiducial
rotation speed in the Tully-Fisher
relation (see eq. 13), and (b)  the Schecter function
exponent, $\alpha$ (see eq. 15).
Within the parameter range,
3 $<$ $\gamma$ $<$ 4, (cf. Giovanelli {\etal} 1997; Sakai {\etal} 2000)
the test statistic distributions  do not change significantly. By contrast,
increasing $\alpha$ above 1.2 reduces {\dlow} significantly owing to the
larger fraction of low mass halos. This would occur if the
damped {\lya} galaxies were drawn from the
luminosity
function measured for the Lyman-break galaxies (Steidel {\etal} 1999),
since in that case $\alpha$ = 1.6 (see $\S$ 2.1).
But overall the improvement of the
model is poor because
$f_{ratio}$ increases to unacceptably high values. The results also change
when we vary $V_{*}$ above 280 {\kms} or below 220 {\kms}. But in both
cases the model does better against some tests and worse against others. In
no case did the TF model
pass all 4 tests
at more than 95 $\%$ nor did the fits of the cross-correlation function
become acceptable.
In that sense the
TF models are in worse agreement with the
data than the CDM models.

\section{SUMMARY AND CONCLUDING REMARKS}

We used
accurate kinematic data acquired for a sample of 35 damped {\lya} systems
to test the standard paradigm of galaxy formation; i.e., the scenario in which
galaxies evolve from the dissipative collapse of virialized gaseous halos onto
rotating disks. The data was presented in the form
of velocity profiles of high ions, intermediate
ions, and low ions in Paper I. In this paper
we considered semi-analytic models,
specifically the MMW models in which centrifugally
supported exponential disks are located at the centers
of dark matter halos drawn from mass distributions
predicted by standard CDM cosmogonies, and where infall of
ionized gas from the halo occurs. We also considered the null
hypothesis that current disk galaxies were in place at
$z$ $>$ 3 (the TF models). We tested the models
with Monte Carlo techniques by comparing distributions
of test statistics generated from observed and model
velocity profiles.
We  utilized eight test statistics: {\dlow}, $f_{edg}$, $f_{mm}$, and $f_{tpk}$
for the low ion profiles (see Paper I), {\dciv} for the C IV profiles, and {\dvm},
$f_{ratio}$, and $\xi(v)$ for comparing low ion and C IV profiles.

First, we discuss the
general implications of our work. As discussed in Paper I
velocity profiles
overlapping in velocity space in such a way that
{\dciv} $\ge$ {\dlow}
are naturally reproduced by scenarios
where
low ion and high ion  kinematic  subsystems are in the same gravitational potential well.
In the collapse scenario the velocity fields of both subsystems scale as {\vc}, yet
more of {\vc} is projected along the line of sight by
gas undergoing radial infall than by gas confined to rotating disks.
This was confirmed by our Monte Carlo simulations of radial
infall of ionized gas clouds onto neutral rotating disks. Indeed in
some cases the infall velocities exceed {\vc} resulting
in {\dciv}/{\dlow} ratios that are too large.
By contrast,
scenarios in which the high ions are embedded in
gaseous outflows (e.g. Nulsen {\etal}  1998)
or any flows
not generated by dark-matter potentials determining
low ion velocities will in general not satisfy these constraints.

Next we discuss specific conclusions arising from this work,
in particular the results of model testing.
Tests of models with the {\em standard} parameters discussed in {$\S$} 4 and
{$\S$} 5 led to the following conclusions:

(1) In the case of the low ion gas none of
measured distributions of {\dlow}, $f_{edg}$, $f_{mm}$, and $f_{tpk}$
were compatible with the predictions of  the CDM cosmogonies at
the 95$\%$ confidence level.
By comparison, the TF model was compatible with the data at
88 $\%$ confidence level.

(2) For the high ion gas we considered only the
{\dciv} distribution. Comparison with the data showed CDM models with the high
column densities predicted by MMW, i.e.,
$\Nperp=\Nperp^{MMW}$,
were in good agreement with the data. CDM models with significantly
lower $\Nperp$ were not as good because they produced overly large {\dciv}.
For the same reasons TF models with high $\Nperp$ were in
better agreement with the data than with low $\Nperp$.

(3) To test model predictions for the relative properties of the high ion and low ion gas
we considered the {\dvm} distribution. The CDM models
were compatible with the data while TF models were not. Apparently
the high rotation-speed TF disks displace the asymmetric low ion profiles
too far from the velocity centroids of the C IV profiles.

(4) Tests of the $f_{ratio}$ distributions showed neither
CDM nor TF model predictions agreed with the data at 95 $\%$ confidence.

(5) Neither the CDM nor TF models predicted C IV versus low ion cross-correlation functions
that were compatible with the data at 99 $\%$ confidence.

We then explored parameter space to determine whether these conclusions
were robust (see $\S$ 6). We varied three crucial parameters: {\vcut}, 
$\Nperp$, and $q_{CIV}$.
This exercise led to the following conclusions:

(6) Figure~\ref{confidencelevels} shows that the NfVR-{$\Lambda$}CDM model
is compatible with the {\dlow}, {\dciv}, {\dvm}, and $f_{ratio}$ tests
at more than 95 $\%$ confidence throughout the small area of the
{\vcut} versus $\Nperp$ plane shown in Figure~\ref{confidencelevels95}.
This is a serious shortcoming since
the NfVR model used for the comparison is ``optimistic'' in that
it predicts a constant velocity rotation curve with maximum
rotation speed {\vrot} = $f_{V}${\vc}. It also
predicts
low impact parameters owing to the restriction $\Nperp \ <  10^{21.2}$
cm$^{-2}$.
In addition, a small decrease in disk thickness would
eliminate consistency throughout parameter space; i.e., the
disks must be thick.

(7) Figure~\ref{confidencelevels_TF} shows the NfVR-TF model
to be compatible with the {\dlow}, {\dciv}, and $f_{ratio}$
tests and incompatible with the {\dvm} test at the 95 $\%$ confidence.

(8) Neither CDM nor TF models produce C IV versus low ion cross correlation
functions that were consistent with the data in the parameter space shown in these
Figures.

What have we learned from the model tests? Because the CDM
models pass 4 out of 5 tests and the TF models pass 3 out of 5 tests,
the CDM models appear to be more plausible. But
to achieve this result
it was necessary  to
adopt a flat rotation curve with {\vrot} = $f_{V}${\vc}.
This is the
maximum rotation speed possible for a model disk, and {\vrot} in realistic protogalactic
disks are probably lower. However, Figure 11 shows that a parameter search for
models with {\vrot}={\vc} reveals no region
in parameter space which is compatible with the 4 kinematic
tests at 95 $\%$ confidence.
Second, the limit $\Nperp < 10^{21.2}$~cm$^{-2}$
indicates that exponential disk models should
predict a steepening of the column-density distribution function at $N$ $>$ 10$^{21.2}$
cm$^{-2}$. This effect is not present in the data (Wolfe {\etal} 1995; Rao \& Turnshek 1999).
Third, the failure of any model to reproduce the C IV versus
low ion $\xi(v)$ indicates significant overlap in velocity space
between the low ion and high ion velocity fields was not achieved. Fourth,
the CDM models predict that most of the damped {\lya} systems occur in low mass halos
where the kinematic state of the ionized gas is highly uncertain (see $\S$ 3.1).

Does this mean that disk-halo models for damped {\lya} systems are ruled out?
We think it is premature to reach this conclusion. Rather
we take these results to mean that if the collapse scenario is correct,
a stronger coupling between the kinematic subsystems
is required. One possibility that comes to mind is for low ions to be associated
with the infalling C IV clouds. This would increase the low ion line-of-sight velocities
and cause smaller differences between the C IV and low ion velocity profiles. But the problem
is there is no evidence for low ions with high ion
kinematics. Another way  to couple the subsystems  is to include the angular momentum
of the halo gas. This
is neglected in the radial infall model.
As the C IV
clouds approach the disk they spin up and experience azimuthal velocity components approaching
{\vrot}. The idea is plausible if the angular momentum vector of the infalling gas is related
to that of the disk, and if clouds near the disk are likely to be detected.
It is encouraging that recent N-body simulations show the angular momenta
of disk and halo to be correlated (Weinberg 2000).
It is also encouraging that
the density of clouds along the line of sight is highest near the disk.
Still, if this idea doesn't work one would be forced to abandon the
disk-halo hypothesis; i.e., one of the standard paradigms of galaxy
formation.

Can the kinematic data be better explained by scenarios other than
infall of ionized gas onto rotating disks of neutral gas?
First, we already discussed problems associated
with outflow models. Second, lacking analytic expressions for
the various quantities, it is not clear whether
numerical
simulations of damped {\lya} systems are compatible with the
kinematics of the ionized gas. However, the density contours in
\cite{haeh98} show the C IV clouds to be within $\sim$ 10 kpc
of the low ion clouds in which case both would be subjected to the
same dark-matter gravitational field.  Because the low ion gas
is not confined to rotating disks, it is not obvious why the
predicted $f_{ratio}$ should exceed the observed lower limit of unity.
To satisfy this constraint one must consider contributions to the
C IV profiles from gas outside the dark-matter halos, perhaps
in the fashion described by \cite{rauch97} for the
C IV QSO absorption lines.
Third, the
scenario described by the semi-analytic modeling of Maller {\etal} (2000)
may provide a good fit to the C IV kinematics. In particular,
both the low ion and C IV kinematics arise from the orbital motions
of mini-halos accreted onto more massive halos with {\vc} $\sim$ 150 {\kms},
implying {\dciv} $\ge$ {\dlow} and that C IV and low ion profiles
are well correlated. The current difficulty with the model is
to physically motivate the very large Mestel disks required to
explain the low ion kinematics.
In any case performing the  tests outlined in this paper
will reveal how robust these models actually are.

\acknowledgements

We wish to thank the following people for valuable discussions:
Robert Benjamin, James Bullock, Eric Gawiser, Chris McKee, Jim Peebles,
and Amos Yahil.  JXP acknowledges support from a Carnegie
Postdoctoral Fellowship. AMW was partially supported by NSF
grant AST0071257.

\appendix

\section{Press-Schecter Theory for ICDM Cosmogonies}

According to Press-Schecter theory the density of bound halos is in the mass
interval $(M, M+dM)$ is given by

\begin{equation}
n(M,z)dM = -2 {\rho \over M}{{\partial F} \over {\partial M}}dM \cmma
\end{equation}

\noindent where $\rho$ is the mean
density of matter and $F$ is the fraction of objects with masses $\le$ M that collapsed by
redshift, $z$. For Gaussian distributed density fields, $F$ is given by

\begin{equation}
F = (1/2) \Bigg [ 1 - {\rm erf} \bigg ( 
{\delta_c \over {\sqrt 2}D(z){\Delta_0(M)}}
\bigg ) \Bigg ]
\cmma
\end{equation}

\noindent where $D(z)$ is the density-contrast growth factor (Peebles 1980) and
$\Delta_{0}(M)$ is the current rms density contrast with mass $M$. While eq. (A2)
is applicable to ACDM models, it does not apply to ICDM models where
the density contrasts are not Gaussian distributed. Rather, we shall
use an analytic fit to results of
numerical computations (Peebles 1998b) which shows that in ICDM, $F$ is given by

\begin{equation}
F = 0.37 {\rm exp}{\bigg [} -0.67 {\bigg (}{{\delta_{c}} \over D(z){\Delta_{0}(M)}}
  \bigg )  
\bigg ]
\perd
\end{equation}

\noindent For  power spectra in which $P(k)$ $\propto$ $k^{n}$, $\Delta_{0}(M)$ $\propto$
$M^{-({{n+3} \over 6})}$. As a result

\begin{eqnarray}
{\rm ICDM}: n(M,z)dM = 0.5 {\rho \over M}
\bigg ( {M \over M_{*}} \bigg )^{({{n+3} \over 6})}
\nonumber \\
\times \; {\rm exp} \Bigg [-0.67 \bigg ({M \over M_{*}} \bigg)^{({{n+3} \over 6})}
\Bigg ] \bigg ( {n+3 \over 6} \bigg ) {dM \over M}
\cmma
\end{eqnarray}

\noindent where $M_{*}$, the mass going non-linear at redshift $z$, is obtained from the
expression, $\Delta_{0}(M_{*})$ = $\delta_{c}/D(z)$ (see White 1997). One finds that

\begin{equation}
M_{*}(z) = M_{8}{\bigg [}{{\Delta_{0}(M_{8})} \over {\delta_{c}}}{\bigg]}^{{6} \over {n+3}}{\bigg [}D(z){\bigg ]}^{{6} \over {n+3}}
\cmma
\end{equation}

\noindent where the mass in a sphere with radius $R$ = 8$h^{-1}$ Mpc, $M_{8}$ = 5.9$\times$
10$^{14}$$\Omega_{M}$$M_{\odot}$. Throughout the paper,
we denote $\Delta_{0}(M_{8})$ by the more conventional symbol,
$\sigma_{8}$. In the case of Gaussian linear density fields applicable to
ACDM models we have

\begin{eqnarray}
{\rm ACDM}: n(M,z)dM = -{\sqrt {2 \over \pi}} {\rho \over M}{{\delta_{c}} \over D{\Delta_{0}}^
{2}}{d\Delta_{0} \over dM} \nonumber \\
\times \;{\rm exp}{\Bigg [}-{{\delta_{c}^{2}} \over 2D^{2}{\Delta_{0}^{2}}}{\Bigg ]}dM
\cmma
\end{eqnarray}

\noindent where in this case $\Delta_{0}(M)$ is determined by the full CDM power spectrum
(Bardeen {\etal} 1986).

\section{Hot Gas in NFW Halos}

Following MM we assume the hot gas is in hydrostatic equilibrium with the dark matter
potential and exhibits adiabatic
temperature and density profiles
out to {\rmin} = min({\rc},{\rv}).
For the NFW rotation curves given in eq. (7)  the density and temperature profiles
of  hot
gas within $r_{min}$ are given by

\begin{eqnarray}
\rho_{h}(r) &= \rho_{h}(r_{min})[1 +0.8Z_{NFW}(r)]^{3/2} \semic
\nonumber \\ 
T_{h}(r) &= T_{h}(r_{min})[1 +0.8Z_{NFW}(r)] \semic \nonumber \\
\quad\quad\quad & 0 < r < r_{min}
\cmma
\end{eqnarray}

\noindent where

\begin{equation}
Z_{NFW}(r) =
{c \over {\delta_{NFW}}}{\Biggl[}{{{\rm ln}(1+x)} \over x}-{{{\rm ln}(1+x_{min})} \over x_{min
}}{\Biggr]}
\semic
x=r/r_{s}
\cmma
\end{equation}

\noindent and

\begin{equation}
{\delta_{NFW}} = [{\rm ln}(1+c)-c/(1+c)]
\perd
\end{equation}

\noindent At $r \ > r_{min}$ the hot gas either follows isothermal profiles when
{\rv} $>$ {\rc} or does not exist when {\rv} $<$ {\rc}. The temperature and density of hot
gas at $r_{min}$ are  given by

\begin{equation}
\ T_{h}(r_{min}) = T_{vir}
\cmma
\end{equation}

\noindent and

\begin{equation}
{\rho_{h}(r_{min})} = {f_{g}V^{2}_{200} \over 4{\pi}Gr^{2}_{s}}{c \over {\delta_{NFW}}}
{1 \over {x_{cool}(1+x_{cool})^{2}}}
\semic x_{cool}=r_{cool}/r_{s}
\perd
\end{equation}

\noindent where $f_{g}$ is the mass fraction in gas.
To solve for $r_{cool}$ we follow MM and let ${\rho_{h}(r_{min})}$ =
5$({\mu}m_{H})^{2}${\vc}$^{2}$/4$\Lambda(T_{vir})t_{M}$
where ${\Lambda(T)}$ is the cooling function. From the last equation we find that
$x_{cool}$ is the root of  the following polynomial:

\begin{equation}
x_{cool}(1+x_{cool})^{2}-(c/{\delta_{NFW}})(R_{cool}/r_{s})^{2}=0
\cmma
\end{equation}

\noindent where $R_{cool}$, the cooling radius of a singular isothermal sphere,
is given by

\begin{equation}
R_{cool} = {\Biggl(}{\Lambda(T_{v})t_{M}m_{d} \over 5{\pi}G{\mu}^{2}m_{H}^{2}}{\Biggr)^{1/2}}
\cmma
\end{equation}

\noindent (see MM). The root of interest is given by $x_{cool} \ \ge$ 1.

\section{Cool Gas}

The cool gas in the halo comprises photoionized clouds formed from
hot phase gas at $r$ $\le$ $r_{min}$ where $r_{min}$ = min({\rc},{\rv}).
Therefore, the mass of gas in the cool phase
is given by

\begin{equation}
M_{cool}= f_{g}M(r_{min}) - M_{h}(r_{min})
\cmma
\end{equation}

\noindent where $M(r_{min})$, the total mass within $r_{min}$, and $M_{h}(r_{min})$, the
mass of hot gas within $r_{min}$, are given by

\begin{equation}
M(r_{min})= { {r_{min}V^{2}_{rot}(r_{min})} \over G}
\cmma
\end{equation}

\noindent  and

\begin{equation}
M_{h}(r_{min})= {\int_{0}^{r_{min}}4{\pi}r^{2}{\rho_{h}(r)}dr}
\perd
\end{equation}

 For NFW halos we use the rotation curve in eq. (7) to relate
$V_{rot}(r_{min})$ to {\vc}, eqs. B1$-$B4  to compute the integral
in eq. C3, and finally eq. B5 to evaluate $\rho_{h}(r_{min})$. The result is

\begin{eqnarray}
M_{cool} = {{f_{g}r_{200}V^{2}_{200}} \over G{\delta_{NFW}}}{\Biggl[}{\rm ln}(1+x_{min})
-{x_{min} \over (1+x_{min})} \nonumber \\
-{{I_{NFW}(x_{min})} \over {(x_{cool})(1+x_{cool})^{2}}} \Biggr]
\semic x=r/r_{s}
\cmma
\end{eqnarray}

\noindent where

\begin{equation}
I_{NFW}(x_{min})= {\int_{0}^{x_{min}}x^{2}[1+0.8Z_{NFW}(x)]dx}
\perd
\end{equation}

\noindent  MM assume the following form for the density of the cool gas:

\begin{equation}
\rho_{c}(r)= {M_{cool} \over {4{\pi}r^{2}V_{c}t_{M}}}
\cmma
\end{equation}

\noindent where $V_{c}$ is a characteristic infall velocity.
Combining the last equation with $\rho_{c}(r)$ = $C_{1}/r^{2}$, we have

\begin{eqnarray}
C_1 = {f_g r_{200} V^2_{200} \over 4\pi GV_c t_M \delta_{NFW}}
\Biggl[ \ln (1+x_{min}) - {x_{min} \over (1+x_{min})} \nonumber \\
- {I_{NFW}(x_{min}) \over (x_{cool})(1+x_{cool})^{2}} \Biggr]
\perd
\end{eqnarray}

Having obtained $C_{1}$ we compute $r_{cross}$ by combining these results with eq. (17).
We find $r_{cross}$ to be the roots of the following equation:

\begin{equation}
r^{2}[1+Z_{NFW}(r)]^{5/2}-{{C_{1}T_{c}} \over {{\rho_{h}(r_{min})}{T_{h}(r_{min})}}}=0
\perd
\end{equation}

We compute the cloud cross-section, $A(r)$, by
combining these results with  eq. 18.  We find that

\begin{equation}
A(r) = {\pi}\left\{
\begin{array}{ll}
{C_{2}}M^{2/3}_{cld}(r)[1+Z_{NFW}(r)]^{-5/3} \ ; r > r_{cross} \\
{C_{3}}M^{2/3}_{cld}(r_{cross}){(r/r_{min})^{4/3}} \ ; r < r_{cross}
\cmma
\end{array}
\right.
\end{equation}

\noindent where

\begin{equation}
C_{2}={\Biggl[}{3T_{c} \over 4{\pi}{\rho_{h}(r_{min})}T_{h}(r_{min})}{\Biggr]^{2/3}},
C_{3}={\Biggl[}{3r_{min}^{2} \over 4{\pi}C_{1}}{\Biggr]^{2/3}}
\perd
\end{equation}

\noindent We  compute the cloud mass by integrating eq. (19) from
$r_{min}$ to $r$ to find $M_{cld}(r)$ at $r \ > r_{cross}$, and assume
$M_{cld}(r)=M_{cld}(r_{cross})$ at  $r \ < r_{cross}$. We find that

\begin{equation}
M_{cld}(r) = \left\{
\begin{array}{ll}
M_{cld}(r_{min}){\Biggl[}1+{{\pi}{\rho_{h}(r_{min})}C_{2}r_{s} \over 3M^{1/3}_{cld}(r_{min})}{
\int_{r/r_{s}}^{r_{min}/r_{s}}[1+0.8Z_{NFW}(x)]^{-1/6}}dx{\Biggr]^{3}} \ ; r > r_{cross} \\
M_{cld}(r_{cross}) \ ; r < r_{cross}
\perd
\end{array}
\right.
\end{equation}

\section{Cloud Kinematics}

The infall velocities of the clouds are computed as follows. At $r \ > r_{cross}$
we compute $V_{r}(r)$ from eq. (22) by using the solutions for $M_{cld}(r)$ given
in Appendix C and noting that the gravitational acceleration is given by
$g(r)$ = ${V^{2}_{rot}(r)}/r$. From eq. (7)  we find

\begin{equation}
g(r) = {V^{2}_{200}}
{1 \over r_{s}}{c \over {\delta_{NFW}}}{1 \over x^{2}} \Bigl [ 
{\rm ln}(1+x)-{x \over 1+x} \Bigr ]  \ ; x=r/r_{s} \
\perd
\end{equation}

At $r \ < r_{cross}$ we consider two scenarios. In the case
of ballistic infall we assume
$M_{cld}(r)$ = const. For clouds with infall velocity $V_{r}(r_{cross})$ at
$r = \ r_{cross}$, the  ballistic solution for NFW halos is given
by

\begin{eqnarray}
V^{bal}_r(x)= \Biggl [  V^2_r (r_{cross}) + 
V^2_{200} {2c \over \delta_{NFW}} \biggl[ { \ln(1+x) \over x} -
\nonumber \\
{\ln(1+x_{cross}) \over x_{cross}} \biggr] \Biggr ]^\ohf
\semic \quad x=r/r_{s} \perd
\end{eqnarray}

\noindent In the case where the kinematics are dominated by random motions we solve the
Jeans equations for $\sigma_{r}(r)$, the radial velocity dispersion of the clouds,
assuming locally isotropic velocity dispersions (Binney \& Tremaine 1987). In this case we 
have

\begin{equation}
{\sigma_{r}(r)^{2}} =
{1 \over {\nu (r)}}{\int_{r}^{\infty}}dx{\nu(x)}g(x)
\cmma
\end{equation}

\noindent where ${\nu(r)}$ is the average density of clouds. Assuming
$\nu(r)$ $\propto$ $\rho_{c}(r)$ we find that

\begin{eqnarray}
{\sigma_{r}^{2}(r)} = {V^{2}_{200}}
{\Biggl(}{c \over 3{\delta_{NFW}}}{\Biggr)}{\Biggl[}{\Biggl(}{1 \over x}-2x^{2}{\Biggr)}{\rm l
n}(1+x)+ \nonumber \\ 2x^{2}{\rm ln}(x)+2x-1{\Biggr]} \ ; 
x=r/r_{s} \
\perd
\end{eqnarray}


\clearpage










\begin{thebibliography}{}

\bibitem[Bardeen et al.\ (1986)]{bard86}
Bardeen, J. M., Bond, J. R., Kaiser, N.,
\& Szalay, A. S. 1986, \apj, 304, 15

\bibitem[Baugh et al. (1998)]{baugh98}
Baugh, C. M., Cole, S., Frenk, C. S.,
\& Lacey, C. G. 1998, \apj, 498, 504

\bibitem[Benjamin \& Danly (1997)]{benj97}
Benjamin, R. A., \& Danly, L. 1997, \apj, 481 764

\bibitem[Binney  \& Tremaine (1987)]{bt87}
Binney, J.,  \& Tremaine, S. 1987, {\em Galactic Dynamics},
(Princeton: University Press), p. 198

\bibitem[Blumenthal et al.\ (1986)]{blum86}
Blumenthal, G. R., Faber, S. M., Flores, R.,
\& Primack, J. R. 1986, \apj, 301, 27


\bibitem[Courteau (1996)]{cord96}
Courteau, S. 1996, ApJS, 103, 363

\bibitem[Courteau (1997)]{cord97}
Courteau, S. 1997, \aj, 114 2402

\bibitem[Dekel \& Silk (1986)]{ds86}
Dekel, A., \& Sily, J. 1986, \apj, 303 39


\bibitem[Fabian (1994)]{fabian94}
Fabian, A. 1994, ARA{\&}A, 32 277

\bibitem[Giovanelli et al.\ (1997)]{giov97}
Giovanelli, R., Haynes, M. P., da Costa, L. N.,
Freudling, W., Salzer, J. J., \& Wegner, G.
\apj, 477, L1

\bibitem[Gonzalez et al.\ (2000)]{gonz00}
Gonzalez, A. H., Williams, K. A., Bullock, J. S.,
Kolatt, T. S., \& Primack, J. R. 2000,
\apj, 528, 145

\bibitem[Haehnelt et al.\ (1998)]{haeh98}
Haehnelt, M. G., Steinmetz, M., \& Rauch, M. 1998 \apj, 495 647

\bibitem[Jedamzik \& Prochaska (1998)]{jedpro98}
Jedamzik, K., \& Prochaska, J. X. 1998 \mnras, 296 430


\bibitem[Kauffmann (1996)]{kau96}
Kauffmann, G. 1996, \mnras, 281, 475

\bibitem[Kepner et al.\ (1997)]{kep97}
Kepner, J. V., Babul, A., Spergel, D. N.
1997, \apj, 487, 61



\bibitem[Lin \& Murray\ (2000)]{lin00}
Lin, D. N. C., \& Murray
2000, astro-ph 0004055

\bibitem[Loveday et al.\ (1999)]{love99}
Loveday, J., Tresse, L., \& Maddox S.,
\mnras, 1999, 310, 281



\bibitem[Lu et al.\  (1996b)]{lu96b}
Lu, L., Sargent, W.L.W., Barlow, T.A., Churchill, C.W., \&
Vogt, S.  1996, \apjsupp,  105, 475

\bibitem[McDonald \& Miralda (1999)]{mcdme99}
McDonald, P. \& Miralda-Escud$\rm \acute e$, 1999,
\apj, 519, 486

\bibitem[Mac Low \& Ferrara\ (1999)]{macf99}
Mac Low, M.-M., \& Ferrara, A.
\apj, 513, 142:1


\bibitem[Mo, Mao, \& White  (1998)]{mmw98}
Mo, H.J., Mao, S., \& White, S. D. M. 1998, \mnras, 295, 336 (MMW)


\bibitem[Mo \& Miralda-Escud$\rm \acute e$ (1996)]{mom96}
Mo, H.J., \& Miralda-Escud$\rm \acute e$, J. 1996, \apj, 469, 589 (MM)

\bibitem[Navarro, Frenk \& White (1997)]{nav97}
Navarro, J.F., Frenk, C.S., \& White, S.D.M. 1997,
\apj, 490, 493 (NFW)

\bibitem[Navarro,\& Steinmetz (1997)]{jnms97}
Navarro, J. F., \& Steinmetz, M.  1997,
\apj, 478, 95

\bibitem[Nulsen et al.\ (1998)]{nulsen98}
Nulsen, P. E. J., Barcons, X., \& Fabian, A. C.
1998,
\mnras, 301, 168

\bibitem[Peebles (1980)]{pbl1980}
Peebles, P.J.E. 1980, {\em The Large-Scale Structure of the
Universe}, (Princeton: Princeton University Press), p. 49

\bibitem[Peebles (1999a)]{pbl99a}
Peebles, P.J.E. 1999a, \apj, 510, 523

\bibitem[Peebles (1999b)]{pbl99b}
Peebles, P.J.E. 1999b, \apj, 510, 531

\bibitem[Prochaska and Wolfe (1997)]{pro97}
Prochaska, J. X. and Wolfe, A. M. 1997, \apj, 487, 73 (PW1)

\bibitem[Prochaska and Wolfe (1998)]{pro98}
Prochaska, J. X. and Wolfe, A. M. 1998, \apj, 507, 113 (PW2)


\bibitem[Prochaska and Wolfe (1999)]{pro99}
Prochaska, J. X. and Wolfe, A. M. 1999, \apjsupp, 121, 369

\bibitem[Prochaska and Wolfe (2000)]{pro00}
Prochaska, J. X. and Wolfe, A. M. 2000, in preparation

\bibitem[Rao \& Turnshek  (1999)]{rt99}
Rao, S. M., \& Turnshek
1999, astro-ph 9909164

\bibitem[Rauch et al.\ (1997)]{rauch97}
Rauch, M., Haehnelt, M.G., \& Steinmetz, M. 1997, \apj, 481, 601

\bibitem[Ostriker \& Rees (1977)]{jpomjr77}
Rees, M. J., \& Ostriker, J. P. 1977,
\mnras, 179, 541


\bibitem[Sakai et al. (2000)]{sakai00}
Sakai, S. {\etal} 2000, \apj, 529, 698

\bibitem[Steidel (1999)]{stei99}
Steidel, C. C., Adelberger, K. L., Giavalisco, M., Dickinson,
M.,  \& Pettini, M. 1998, \apj, 519, 1

\bibitem[Thoul \& Weinberg (1995)]{tw95}
Thoul, A. A., \& Weinberg, D. H. 1995,
\apj, 442, 480


\bibitem[White  (1996)]{wht96}
White, S.D.M., 1996, in {\em Cosmology \& Large Scale Structure}, ed.
R. Schaeffer, J. Silk, M. Spiro, \& J. Zinn-Justin, (Amsterdam:Elsevier),
p. 349


\bibitem[White \& Frenk (1991)]{white91}
White, S.D.M., \& Frenk, C.S. 1991, \apj, 379, 52


\bibitem[Weinberg \ (2000)]{wein00}
Weinberg, D., 2000, private communication


\bibitem[Wolfe et al.\ (1995)]{wol95a}
Wolfe, A. M., Lanzetta, K. M., Foltz, C. B., \&
Chaffee, F. H. 1995, \apj, 454, 698

\bibitem[Wolfe \ (2000)]{wol20}
Wolfe, A. M.,  \& Prochaska, J. X., 2000,
\apj, submitted (Paper I)



\end{thebibliography}
\end{document}